\documentclass[useAMS,usegraphicx]{mn2e}

\title[The WFCAM Science Archive]{The WFCAM Science Archive}

\author[Hambly et al.]
{N.C.~Hambly,$^1$ R.S.~Collins,$^1$
N.J.G.~Cross,$^1$ R.G.~Mann,$^1$ M.A.~Read,$^1$ \newauthor
E.T.W.~Sutorius,$^1$ 
I.A.~Bond,$^2$ J.~Bryant,$^1$ J.P.~Emerson,$^3$ A.~Lawrence,$^{1}$ \newauthor
J.M.~Stewart,$^4$ P.M.~Williams,$^1$
A.~Adamson,$^{5}$ S.~Dye,$^{6}$ P.~Hirst,$^{5,7}$ \newauthor
and S.J.~Warren$^{8}$\\
$^1$ Scottish Universities Physics Alliance (SUPA),
     Institute for Astronomy, School of Physics, University of Edinburgh, \\
   ~~Royal Observatory, Blackford Hill, Edinburgh EH9 3HJ \\
$^2$ Institute of Information and Mathematical Sciences, 
     Massey University at Albany, Auckland, New Zealand\\
$^3$ Astronomy Unit, School of Mathematical Science, 
     Queen Mary University of London, Mile End Road, London E1 4NS\\
$^4$ United Kingdom Astronomy Technology Centre, Royal Observatory, 
     Blackford Hill, Edinburgh EH9 3HJ\\
$^5$ Joint Astronomy Centre, 660 N.~A`oh\=ok\=u Place, University Park, Hilo, 
     Hawaii 96720, USA\\
$^6$ School of Physics and Astronomy, Cardiff University, 5 The Parade, 
     Cardiff CF24 3YB, Wales, UK\\
$^7$ Gemini Observatory, 670 N.~A`oh\=ok\=u Place, University Park, Hilo, 
     Hawaii 96720, USA\\
$^8$ Blackett Laboratory, Imperial College of Science Technology
     and Medicine, Prince Consort Rd, London SW7 2AZ}

\begin{document}

\date{Accepted XXXX. Received XXXX ; in original form XXX }

\pagerange{\pageref{firstpage}--\pageref{lastpage}} \pubyear{2005}

\maketitle

\label{firstpage}

%++++++++++++++ ABSTRACT +++++++++++++++++++++++++++++
\begin{abstract}

We describe the WFCAM Science Archive (WSA), which is the primary point
of access for users of data from the wide--field infrared camera WFCAM
on the United Kingdom Infrared Telescope (UKIRT), especially 
science catalogue products from the UKIRT 
Infrared Deep Sky Survey (UKIDSS). We describe the database design with emphasis
on those aspects of the system that enable users to fully exploit the
survey datasets in a variety of different ways. We give details of
the database--driven curation applications that take data from the standard
nightly pipeline--processed and calibrated files for the production
of science--ready survey datasets. We describe the
fundamentals of querying relational databases 
with a set of astronomy usage examples, and illustrate the
results.

\end{abstract}


%++++++++++ BODY OF PAPER ++++++++++++++++++++++++++

\section{Introduction}  \label{intro}

The term `science archive' is first seen in the astronomy literature
in Barrett~(1993) which describes the High Energy Astrophysics Science
Archive Research Centre (HEASARC). This system is much more than a
simple repository of data -- HEASARC provides an online resource to
enable scientific exploitation of high--energy astronomy missions
via provision of science data, software, analysis tools and
descriptive information. For example, the data holdings in the
HEASARC amount to many terabytes (TB; 1~TB=10$^{12}$~bytes) so
wholesale download is impractical; recognising this, a {\em server--side}
analysis facility (i.e.~a facility co-located with the data and hence
remote to the typical user) is provided to enable large--scale 
processing given an arbitrary astronomical usage scenario. In this
way, data download is limited to user--defined subsets, sometimes
processed in a manner specified by the user at access time.

The advent of the large Schmidt photographic plate digitisation
programmes (Hambly et al.~2001a and references therein) and infrared
surveys such as DENIS (Epchtein et al.~1994) and 2MASS 
(Kleinmann et al.~1994) presented similar challenges for 
ground--based missions. Data distribution for the digitised Schmidt
surveys was originally done on removable, permanent storage media
(`compact disc' read--only memory), but this became
impractical so online services rapidly developed
for these also. However, it is probably fair to say that it was
with the challenges posed by the Sloan Digital Sky Survey (SDSS;
York et al.~2000) that the ground--based astronomical science
archive became fully developed (Gray et al.~2002). The first SDSS
(the so--called Sloan Legacy Survey)
is now complete, and $\sim2\times10^{8}$ sources have been measured
and characterised, producing a catalogue of several~TB in size with
associated imaging data and $\sim10^6$ spectra amounting to a total
volume of $\sim10$~TB (Adelman--McCarthy et al.~2007 and references
therein); the state--of--the--art SDSS science archive is described 
in Thakar et al.~(2003a).

The challenges and opportunities presented by the current generation
of ground--based infrared surveys were noted by Lawrence et al.~(2002).
In particular, they cited the advent of a new wide--field camera
for the 4-m United Kingdom Infrared Telescope (WFCAM for UKIRT; 
Casali et al.~2007)
and the even greater challenges posed by the new dedicated 4-m telescope
for infrared surveys VISTA (Emerson~2001). These ambitious survey
missions gave rise to a systems--engineered data management project,
the VISTA Data Flow System (VDFS; Emerson et al.~2004) which included
provision of pipeline processing and science archiving for WFCAM and
VISTA data. Here, we concentrate on the first VDFS science archive
known as the WFCAM Science Archive (WSA). From the outset, the design
of the WSA has been science--driven with the main science
stake--holders being users of the UKIRT Infrared Deep Sky Survey (UKIDSS;
e.g.~Warren~2002).

This paper is one of a set of five which
provide the reference technical documentation for UKIDSS, although it
is of direct relevance to any user of the WFCAM Science Archive.
The other four papers in the series describe the infrared survey instrument
itself (WFCAM; Casali et al.~2007); the WFCAM photometric system (Hewett et 
al.~2006); the UKIDSS surveys (Lawrence et al.~2007); and the pipeline
processing system (Irwin et al.~2007).

This paper is arranged as follows. In Section~\ref{design}, we
describe the design of the WSA, concentrating on the development
of the science requirements into data models (i.e.~the database
design) as presented to the end--user at access time. In
Section~\ref{browser} we discuss various detailed implementation
issues that in particular inform the user as to how science--ready
survey catalogues are generated from the standard 
flat file products processed by the nightly pipeline.
Section~\ref{queries} then
goes on to discuss some illustrative science examples by 
concentrating on the expression of certain specific science usage
modes in Structured Query Language (SQL), the {\em lingua franca} of
relational database users. Following the usual conclusion,
acknowledgements and bibliography we present as appendices
some supplementary information to aid first--time users of the
WSA.

\section{Design} \label{design}

In this paper, we concentrate on those aspects of the design that
are relevant to the end user, assumed to be an astronomer interested
in exploiting the archive for the purposes of scientific research.
Further background information, and in particular technical details of the 
Information Technology aspects of the overall VISTA Data Flow System 
can be obtained from a set 
of papers appearing in recent volumes of the Astronomical Data Analysis
and Software Systems (ADASS) and the International Society for Optical
Engineering (SPIE) publications series -- see 
Hambly et al.~(2004a), Collins et al.~(2006), Emerson et 
al.~(2006) and Cross et al.~(2007).
The design of the WSA is based, in part, on that of
the science archive system for the SDSS
(Thakar et al.~2003a and references therein).
In particular, we have made extensive use of the relational design
philosophy of the SDSS science archive, and have implemented
some of the associated software modules (e.g.~that for the computation
and use of Hierarchical Triangular Mesh indexing of spherical
coordinates -- see Kunszt et al.~2000). Scalability of the design to
terabyte data volumes was prototyped using our own existing legacy
Schmidt survey dataset, the SuperCOSMOS Sky Survey (Hambly et al.~2001a
and references therein). The resulting prototype science archive system,
the SuperCOSMOS Science Archive (SSA) is described in 
Hambly et al.~(2004b) and provides an illustration of the contrast in
end--user experience of an old--style survey interface (as described in
Hambly et al.~2001) and the new. Note that extensive
technical design documentation for the WSA is maintained 
online\footnote{\tt http://surveys.roe.ac.uk/wsa/pubs.html}.

The following Sections provide more information on the design to a
level of detail that will enable a general user of the WSA to understand
and to get the highest possible return out of the system.

\subsection{Background}

The WSA is a system designed to store, curate and 
serve all observations made by WFCAM, which is described in detail 
in Casali et al.~(2007). The infrared active
part of the focal plane consists of four $2048\times2048$ detectors with
plate scale 0.4\arcsec\ pix$^{-1}$ arranged in a square pattern and
spaced by 94\% of the detector width (e.g.~Casali et al.~2007, Figure~2).
Hence, a sequence of four pointings is required to produce contiguous
areal coverage of 0.78~sq.~deg.~(this is sometimes called a {\em tile});
however the unit of WSA curation (e.g.~frame association for source merging
-- see later) is based around images of the size of
one detector (known as a {\em detector frame}). Such an image is usually
the result of stacking of a set of {\em dithered} and/or {\em microstepped}
individual exposures (known as normal frames in the VDFS). Dithering (also
known as {\em jittering}) is typically executed in step patterns of several
arcseconds about a base position to allow for the removal of poor
quality pixels at the processing stage. Microstepping, on the other hand,
is sometimes used to recover full PSF sampling as the image quality
delivered by WFCAM/UKIRT often can be better than the Nyquist limit of
$\sim0.8$\arcsec\ given the 0.4\arcsec\ WFCAM pixels. 
WFCAM instrument performance is concisely summarised in Casali et 
al.~(2007), Table 3: e.g.~median (best) image quality is 0.7\arcsec\
(0.55\arcsec) at zenith at K band.

Observing time with WFCAM on UKIRT
is divided between large scale surveys (i.e.~UKIDSS and the recently
instigated `campaigns'), smaller PI--led projects (awarded time via a
telescope time allocation group), `service' mode observations for very
small projects requiring only a few hours of time, and special projects
like observatory/survey infrastructure (calibration) and director's
discretionary time projects. Data from all these is tracked in the WSA,
but the design is dictated primarily by the largest surveys, i.e.~UKIDSS,
which is described in detail in Lawrence et al.~(2007). 

Briefly, UKIDSS consists of a hierarchy of five surveys that trade
depth versus area to cover a multitude of science goals. The Large
Area Survey (LAS) aims to cover $\sim4000$~sq.~deg.~in four infrared
passbands to depths Y~$\sim20.3$, J~$\sim19.8$, H~$\sim18.6$ and
K~$\sim18.2$ with two epochs of coverage at~J. The Galactic Plane
Survey (GPS) aims to cover $\sim1900$~sq.~deg.~to depths J~$\sim19.9$,
H~$\sim19.0$ and K~$\sim19.0$ with two (originally three) 
epochs of coverage at K and
some coverage at narrow--band H2. The Galactic Clusters Survey (GCS)
will survey ten open--cluster/star--formation regions to a total of
$\sim1000$~sq.~deg.~to depths Z~$\sim20.4$, Y~$\sim20.3$, J~$\sim19.5$,
H~$\sim18.6$ and K~$\sim18.6$ with two epochs of coverage at~K. The
Deep eXtragalactic Survey (DXS) aims to survey four selected areas to a
total of $\sim35$~sq.~deg.~to depths J~$\sim22.3$, H~$\sim21.8$ and
K~$\sim20.8$. Finally, the Ultra Deep Survey (UDS) aims to survey
$\sim0.8$~sq.~deg.~to depths J~$\sim24.8$, H~$\sim23.8$ and K~$\sim22.8$.
UKIDSS LAS (J), GPS (JHK), GCS (K) and DXS (JK) employ $2\times2$ 
microstepping (yielding 0.2\arcsec\ samples) while the UDS employs 
$3\times3$ microstepping in all filters (yielding 0.13\arcsec\ samples).
In the VDFS, an image resulting from interleaving microstepped frames is
known as a {\em leav} frame while an image resulting from stacking a set of
dithered exposures is known as a {\em stack}. An interleaved, stacked
image is known as a {\em leavstack} frame -- for many more details of
VDFS pipeline processing, see Irwin et al.~(2007).
Survey data quality obtained in practice is summarised in UKIDSS
data release papers (e.g.~Dye et al.~2006; Warren et al.~2007a).
Median seeing at Data Release~1 was $\sim0.83$\arcsec; uniformity
of photometric calibration as estimated via field--to--field scatter
was between~0.02 and~0.03 mag in Y--J, J--H and H--K; mean stellar
ellipticity was $\sim0.08$. Observing strategies for UKIDSS are
discussed extensively in Lawrence et al.~(2007). Tiling the wide,
shallow surveys, especially at high Galactic latitude, is dictated
largely by the availability of suitable guide stars (V~$<17$; Casali
et al.~2007). This results in varying degrees of frame overlap and
non--uniform tiling. The WSA copes with this via a data--driven
source merging philosophy, and a flexible seaming algorithm for
the production of interim catalogue products during the
7~year UKIDSS observing campaign, as is required to maximise timely
scientific exploitation. Furthermore, a requirement exists for
associating multiple--epoch visits of the same field, in
addition to merely associating different passband visits.
Again, a database--driven application ensures that
sensible frame associations are made in the presence of
incomplete datasets when intermediate releases are required
before full survey completion.

In WSA parlance, UKIDSS as a whole is referred to as a {\em survey}
while the LAS, GPS, DXS etc.~are known as {\em programmes} (the
rest, including PI--led programmes, are known as 
{\em non--survey programmes}). For the
purposes of book--keeping at the observatory, observing is broken up
into chunks known as {\em projects} which have a unique name that
may include a Semester identification (e.g.~u/07a/32 for non--survey
PI--led programme no.~32 in Semester~07A; u/ukidss/gcs5 for UKIDSS
GCS project observing set no.~5). The various survey and non--survey 
processed datasets stored and served in the WSA have proprietary periods 
ranging from~12 months for non--survey programmes to~18 months for
the larger campaigns and surveys. These periods run from the time
at which the processed data are made available to the respective
proprietors rather than individual frame observing dates. Note that
UKIDSS is proprietary to astronomers in the European Southern
Observatory member states, while campaign and non--survey datasets
are proprietary to the respective PIs and their named collaborators.

\subsection{Archive requirements}

A set of top--level general requirements was established early in the
history of the WSA 
project\footnote{{\tt http://www.jach.hawaii.edu/UKIRT/management/wds/}
{\tt requirements/wfarcrq.html}}.
Briefly, requirements were specified in the following broad
categories: i) top--level; ii) general contents and functions; iii)
detailed functional requirements; and iv) security. Examples include
i) broad--brush requirements concerning  flexibility, scalability,
ease--of--use and scope, e.g.~the WSA is required to hold all
pipeline--processed WFCAM data, not just that belonging to survey
programmes (UKIDSS); ii) minimal requirements concerning contents
and functionality, e.g.~contents to include pixel, catalogue and
associated metadata, along with calibration data; iii) a set of
detailed functional requirements from the point of view of the
end--user, e.g.~searching and visualisation functionality required
in the user interface; and finally iv) security rules concerning
protection of the data itself, its integrity and any proprietary
rights thereof.

In order to progress the design of the WSA from the top--level
generalities summarised above, we followed a rational process
similar to that employed in the design of the science archive for the
SDSS (Thakar et al.~2003a), viz. the development of a set of
questions and usage modes that one would ask or require of the
archive to fulfill the functional requirements previously
identified. This may seem somewhat {\em ad hoc} compared with
a standard, `unified rational process' (such as is encapsulated
in Unified Modelling Language design, e.g.~Gaessler et al.~2004
and references therein) but it has been
successfully employed in the past (not least in the SDSS science
archive design) and is rather powerful, despite its relative
simplicity. We developed a set of 20 curation usage modes for the
WSA and a set of 20 end--user usage modes in collaboration with
the UKIDSS user community (see Appendix~\ref{usages}).
These were then analysed along with
the original top--level requirements  to produce a requirements
analysis document to inform the detailed design described below.
The design documents are all available 
online\footnote{\tt http:://surveys.roe.ac.uk/wsa/pubs.html}.

\subsection{Design fundamentals}

The WSA receives processed data from the pipeline component of the overall
data flow system in the form of FITS (Hanisch et al.~2001) image and catalogue
binary table files (Irwin et al.~2007). No raw pixel data are held in the
WSA. Processed data consists of instrumentally--corrected WFCAM frames,
associated descriptive and calibration data (including confidence
frames and calibration images,
e.g.~darks and flats) and single--passband detection lists derived from
the science frames. Calibration information also includes astrometric
and photometric coefficients derived using the 2MASS 
(Skrutskie et al.~2006) point source
catalogue as a reference. Metadata, meaning in this context those data
that describe the imaging observations and processing thereof 
(e.g.~observing dates/times, filters, instrument state, weather conditions,
processing steps, etc.), are defined by a set of descriptor 
keywords agreed between the archive and pipeline centres, and include all
information propagated from the instrument and observatory, along with
additional keywords that describe the processing applied to each image
and catalogue in the pipeline. Single passband detection catalogues for
each science image have a standard set of 80 photometric, astrometric and
morphological attributes along with error estimates, and a variety of
summary quality control measures (e.g.~seeing, average point source
ellipticity etc.). For many more details, see Irwin et al.~(2007).

The design of the WSA was based, from the outset, on a classical
client--server architecture employing a third--party back--end 
commercial database
management system (DBMS). This followed similar but earlier developments
for the SDSS science archive, and reflects the great flexibility of
such a system from the point of view of both applications development
and end--user querying. Furthermore, although originally built 
(Szalay et al.~2000) on an `object oriented' database,\footnote{a 
system that presents database objects (tables, rows, columns,
constraints etc.) as programming language `objects' (i.e.~entities
encapsulating both data and programming functionality) to client
applications} 
issues with performance and ease of use by the end user led to a switch 
to a {\em relational} database management system (RDBMS; a system that
presents data as a group of related tabular data sets) in that project 
(Thakar et al.~2003b) and the WSA has been
based on the relational model from the start. This brings many advantages
for astronomy applications (indeed, for applications in any scientific 
discipline) where
related sets of tabular information are familiar. Such advantages will
be illustrated below; at this point we emphasise a few fundamental
aspects of the relational design.

\subsubsection{Default values and `not null'}
\label{notNull}

As always in database design, a decision has to be made as to how to
deal with the situation when no measurement is available to populate
a particular field of a given row. For example, it may be that the
data model (see later) requires that a merged source table has columns
for infrared colours (J--H). What happens when~H, or~J, or even both
are unavailable for that particular source (perhaps the images in 
these filters have not been taken yet, but we require to allow users
access to the data that {\em do} exist for this source -- e.g.\ 
observations in other filters)? This particular attribute, (J--H),
could be set to a specified default value (an appropriately 
out--of--range but nonetheless real number, say $-0.999999\times10^9$)
or it can be allowed to be undefined (`null') in the RDBMS. One of
the (many) problems with null values is that they complicate
querying of the database: it is easier and clearer to ask ``give me
all the objects with (J--H) in the range~0.5 to~1.0'' than it is to
ask the same question with the additional predicate ``and (J--H)
is not null'' (necessary because the RDBMS returns null values in 
results sets as a standard data type to be handled by querying applications).
By judicious choice of default values, we can force 
exclusion of those rows where no measurement is available 
in an explicit and clear manner (in this
case because the default value is outwith the range of a typical
colour selection) thus simplifying querying applications.
In this simple example this may seem rather unimportant but in
more complicated situations the use of default values can greatly
simplify querying applications and, as we describe later, the WSA
philosophy is to expose the full power of the RDBMS to the end
user for complete flexibility in querying. The WSA employs the
default values as specified in Table~\ref{defaults} for the
various data types listed, and does not allow null
attributes in any column of any table.

\subsubsection{Physical units}

Physical quantities in the WSA are stored in SI units wherever possible.
Astronomical convention dictates the usual standards for many 
astrophysical quantities; a conventional magnitude scale on the natural
WFCAM system (Hewett et al.~2006) is employed for calibrated fluxes.
All timestamps employed in the data flow system, including the science
archive, are `Universal Time Coordinate' (UTC) date/times. Spherical
coordinates are stored in equatorial (J2000.0 equinox), Galactic and
SDSS $(\lambda,\eta)$ coordinates (Stoughton et al.~2002)
for ease of querying in different
systems, and all angles (RA, Dec etc.) are stored in units of decimal
degrees apart from a small number of image attributes that map directly
to FITS keywords delivered by the pipeline. Equatorial coordinates at
equinox J2000.0 are labelled with a 20--level Hierarchical Triangular
Mesh index (Kunszt et al.~2000) to make spatially limited queries
efficient.

\subsubsection{Miscellaneous fundamentals}

Pixel data are stored as flat files in the WSA system, rather than as
`binary large objects' in DBMS tables. This is so that high data volume
usages (i.e. those requiring access to pixel data) that are not 
time--critical will not impact catalogue querying, where more `real--time'
performance is required for data exploration and interaction. However,
pixel file names and the pixel metadata are tracked in tables within
the DBMS so that the image descriptors can be browsed and queried in the
same way as, or in conjunction with, catalogue data.

The WSA is organised as a self--describing database. This means that 
{\em curation information}, i.e. information pertaining to 
database--driven activities (for both invocation and results logging) in
preparation of science--ready data products (see later) is contained
in the database, along with science data. For example, the
requirements for source merging for a survey programme (the filter
selection and the number of passes in each filter, the source
pairing criterion, etc.) are stored in database tables to drive the
relevant curation activity and to inform users of the procedure.

\begin{table}
\caption[]{Default values for various data types in the WSA database.}
\label{defaults}
\begin{tabular}{ll}
\hline
Default value          & Data type \\
\hline
$-0.9999995\times10^9$ & Floating point (single/double precision)\\
$-99999999$            & Integer (4-- and 8--byte)\\
$-9999$                & Integer (2--byte)\\
NONE                   & Character\\
9999--Dec--31            & Date--times\\
\hline
\end{tabular}
\end{table}

\subsection{The WSA relational model}
\label{datamodel}

%In order to service the requirements outlined previously, we decided to
%adopt a single data model to service both curation operations (e.g.\
%data ingest, source merging etc.) and end user usage modes. Hence,
%the database design (i.e.\ the arrangement of data in the archive)
%is of direct relevance to the end user, and forms a fundamental aspect
%of the user interface to the data. In this Section, we describe
%the conceptual data model of the WSA.

A good design for a relational database captures the structure
inherent in the data to be stored, thereby aiding curation 
operations and end--user query modes, as these are both likely to
reflect that structure. In conventional relational design, this 
structure is captured in an entity--relational model (ERM), in which 
a collection of related data is represented by an {\em entity\/} 
and entities have relationships between them, which can be mandatory
or optional and can have one of three cardinalities (one--to--one,
one--to--many or many--to--many).  

\begin{figure}
\includegraphics[width=84mm,angle=0]{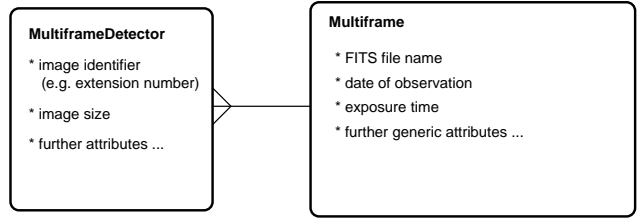}
\caption{A simple `entity--relationship' model (ERM) showing in
schematic form the relationship between the generic attributes of
a multi--extension FITS image (a \texttt{Multiframe} in WSA
parlance) and the particular attributes of each constituent image
(\texttt{MultiframeDetector}) of that multiple image container file.
The one--to--many (in fact one--to--four in the case of WFCAM)
relationship between these two entities is represented by the
`crows foot' connecting the boxes (see text for a more detailed
explanation).}
\label{simpleERM}
\end{figure}

To illustrate this, consider a processed WFCAM image file, as 
delivered by the pipeline (Irwin
et al.~2007). Such a multi--extension FITS (MEF) file consisting of
a primary header--data unit
with generic descriptive keywords (observation
date/time, filename, telescope/instrument parameters etc.) and a set
of extensions containing the images and corresponding descriptive
data of individual detectors can be represented in relational terms
as in Figure~\ref{simpleERM}. Here, we identify entities
\verb+Multiframe+ and \verb+MultiframeDetector+ and a one--to--many
relationship between them, each entity containing attributes that
describe it. 
%This is known as an entity--relationship model (ERM)
%and is a convenient way to describe the arrangement of data in a
%relational database. Each entity maps directly to a table
%in the database and attributes represent columns, or fields in each
%row of those tables. 
A particularly important point to note
here is that the arrangement of data as represented in 
Figure~\ref{simpleERM} is {\em normalised} in the sense that we
do not duplicate attributes in entity \verb+MultiframeDetector+
that pertain to a set of individual extension frames in each
\verb+Multiframe+ -- e.g.\ we could represent the data using a
single entity where each set of detector frames 
(four in the case of WFCAM MEF file of a typical observation)
is described by the generic attributes in entity
\verb+Multiframe+ in addition to the specific attributes pertaining
to each. Clearly, in terms of storage it is more efficient to have
one record of the generic attributes of each set of detector frames,
and link each \verb+MultiframeDetector+ to its parent \verb+Multiframe+
using a label and a reference in the RDBMS. Note that there is no
requirement here for every \verb+Multiframe+ to have exactly
four detector frames. A mosaiced image product can be
equally well described by this data model -- there will simply be
a single extension representing the whole image, and the mosaic
\verb+Multiframe+ will simply have one related row in 
\verb+MultiframeDetector+.

In designing the WSA
relational model, normalisation has been used except in a small
number of cases where it makes sense to {\em denormalise} and
duplicate some attributes for ease of use and better
performance at query time; this is
illustrated later, along with example usage modes requiring to
query a set of normalised tables (`join' queries).
The principle of normalisation complicates the data model for
the novice user, but it is extremely important when designing
a system that must scale to very large data volumes. 

Experience with the SDSS has shown 
that scientifically realistic queries often require inclusion of
constraints on metadata parameters and selection of rows on the 
basis of their provenance (e.g.\ properties of their parent images). 
To do this requires the user to know
the basic structure of the database, so, in the remainder of this 
Section we describe the principal contents of the WSA in terms of ERMs,
at a level which will enable users to define the queries they need
to run to do their science. 

\subsubsection{Image data}

\begin{figure*}
\includegraphics[width=140mm,angle=0]{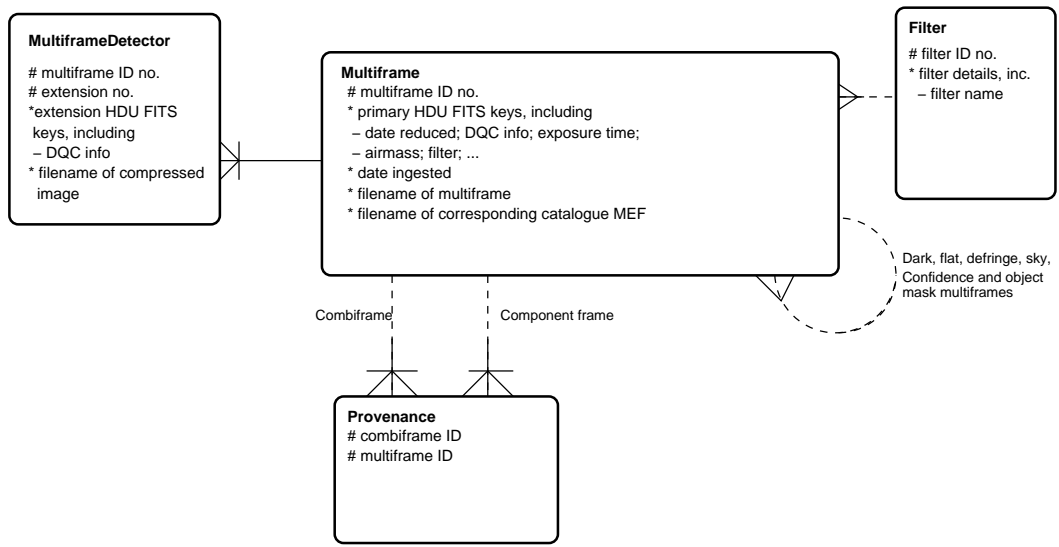}
\caption{Relational model for image data in the WSA. Each box
represents a table in the database; the lists of attributes in
each are for illustration only, and are not intended to be
complete. The `crows feet' illustrate one--to--many relationships
between data entries in each entity; dotted lines indicate optional
as opposed to mandatory relationships (see text for further details).}
\label{imageERM}
\end{figure*}

As noted previously, image metadata are tracked in the WSA database,
although the image pixel data themselves are not ingested into the
RDBMS -- they are stored as flat files on disk. In Figure~\ref{imageERM}
we show the ERM for pixel data in the WSA. Each entity box represents a
database table, and one--to--many relationships between the tables
are shown, as before. Note that some relationships are mandatory
whereas some are optional. An example of a mandatory relationship is
that every \verb+Multiframe+ has one or more \verb+MultiframeDetector+s
(not unreasonably, since a MEF devoid  of any detector frames is not
particularly useful). An example of an optional relationship,
denoted by a dashed line on the side where the relationship is 
optional, is that every \verb+Filter+ {\em might} have one or more
\verb+Multiframe+s (again, not unreasonable since there may be
unused filters present in WFCAM) and yet every \verb+Multiframe+
has to have one associated filter record only. In this case, 
the mandatory relationship implies that there must {\em always} 
be a link
between the \verb+Multiframe+ and \verb+Filter+ tables, even if
that link points to a blank filter record, or if the filter 
keyword in a given \verb+Multiframe+ was unavailable for some
reason, then the link will take a default value (see previously).
However, to maintain referential integrity in the database there
will need to be a default {\em row} in table \verb+Filter+ that
can be referenced by the default link. This situation can occur
in any part of the WSA data model where a mandatory relationship
exists between two tables.

Another useful feature of these ERM schematics is the indication of
a unique identifier using the `\verb+#+' sign in the attribute list
(a convention in entity--relationship modelling). Unique
identifiers (UIDs) are, of course,
key to efficient operation in a DBMS -- without them, a table is
simply a heap of data in which a specific row cannot be found
easily. With a UID, on the other hand, every row of a table is
uniquely labelled and can be located quickly, especially if the
table data are sorted on that attribute (as is generally the case).
Note that barred relationships in Figure~\ref{imageERM} indicate
where the combination of UIDs in both tables linked by the
relationship are used, in the table on the barred side, as a
combined UID. In the case of \verb+Multiframe+ and
\verb+MultiframeDetector+, for example, the UID in the former is
simply a running number assigned on ingest, while in the latter 
the UID is a combination of the parent \verb+Multiframe+ UID plus
the extension number -- in this way, every \verb+MultiframeDetector+
is uniquely identified (it is conventional in ERMs to omit as
`\verb+#+' UID attributes those UIDs from a related table, but we have
explicitly noted them for clarity).

Other types of relationship are shown in Figure~\ref{imageERM},
and they illustrate how the WSA tracks the processing history, or
{\em provenance}, of each processed image. 
%Entities \verb+DifferenceImage+,
%\verb+StackImage+ and \verb+MosaicImage+ are included to track any
%attributes of images of each of those types that are not already 
%contained in entity \verb+Multiframe+, hence each \verb+Multiframe+
%{\em may} have one associated record in the appropriate table if it
%is of that specific image product type; the UID in each case is
%simply the UID of the \verb+Multiframe+ (as indicated by the barred
%relationship). 
Entity \verb+Provenance+ tracks the ancestor images of
any image in the WSA that is the result of a combinatorial
process on other images also tracked in the archive; hence for a
\verb+Multiframe+ composed of~$N$ other \verb+Multiframe+s (e.g.~a
stack of individual dithered \verb+Multiframe+s)
this would contain $N$ records, each consisting of the UID of the final
stack product (the attribute labelled as \verb+combiframeID+) along
with one of each of the $N$ separate constituent \verb+Multiframe+
UIDs; the other optional one--to--many relationship between
entities \verb+Provenance+ and \verb+Multiframe+ indicates that 
every component frame recorded in the former must be present in
the latter, while every \verb+Multiframe+ {\em may} be included as
a constituent frame in one or more combined frame products. Finally,
there is an optional self--referencing relationship indicated in the
lower right--hand corner of entity \verb+Multiframe+. This indicates
that each \verb+Multiframe+ {\em may} be a pixel value correction
frame used in the processing of one or more science \verb+Multiframe+s
(there is an attribute to distinguish between different 
\verb+Multiframe+ types); conversely, each \verb+Multiframe+ {\em may}
have been processed using one or more of each of the correction frame types
dark, flat, sky etc. The relationship is optional on both sides since,
for example, a flat will not itself be calibrated against a flat;
moreover every single calibration frame that is propagated through
the system may not get used in the processing of any science frames.

\subsubsection{General catalogue data model}
\label{genericCat}

\begin{figure*}
\includegraphics[width=120mm,angle=0]{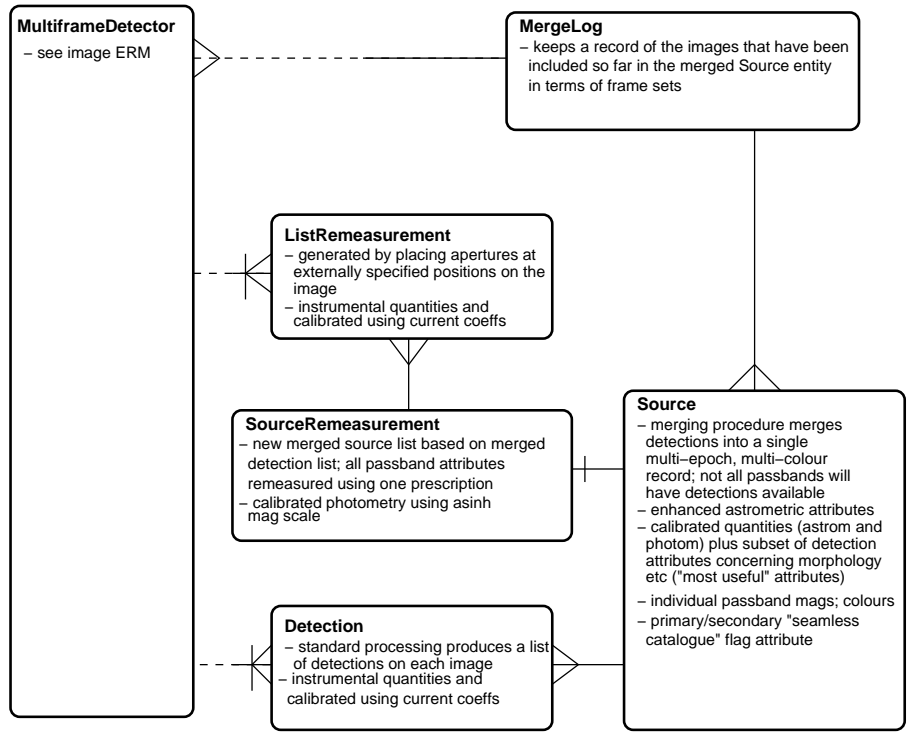}
\caption{Generalised relational model for catalogue data in the WSA
(see text).}
\label{catERM}
\end{figure*}

In Figure~\ref{catERM} we show a generalised ERM for catalogue data in 
the WSA. A set of five entities are identified that link with each other
and with entity \verb+MultiframeDetector+ (see Figure~\ref{imageERM}) as
shown. Briefly, standard 80--parameter detection lists from science
images delivered by the pipeline (Irwin et al.~2007) are tracked in
entity \verb+Detection+; hence every \verb+MultiframeDetector+
{\em may} give rise to one or more \verb+Detection+s with a UID that
includes the UID of the former. End--user science requirements, however,
specify that most science applications need a merged, multi--colour,
multi--epoch source list for convenience, so this data model includes
an entity \verb+Source+ to track merged source records produced by a
standard {\em curation} procedure (see later). Each \verb+Source+
is always made up of one or more individual passband \verb+Detection+s.
The source merging procedure operates on sets of \verb+MultiframeDetector+s
where a frame set comprises detector frames taken at the same position
but in different filters and/or at different times.
These {\em frame sets} are tracked by entity \verb+MergeLog+ where
every \verb+MergeLog+ frame set always consists of one or more
\verb+MultiframeDetector+s while an individual frame in the latter 
may or may not be a member of a frame set -- non--science
frames would not be included in frame sets, for example. The final
two entities in Figure~\ref{catERM} are included to track enhanced
catalogue extraction data from a process known colloquially as
{\em list--driven remeasurement}. Standard pipeline processing treats
each science image separately and extracts sources using a set of
standard apertures and adaptive profile models applied at positions
having detections above a sky noise--dependent threshold as described
in Irwin et al.~(2007). In the list--driven remeasurement
scenario, a frame set is reanalysed for photometric attributes amongst
all individual frames in the set using a master list of sources that
are present in the field and a single set of apertures and models to
yield photometric attributes consistently measured across the frame
set. In this way, attributes such as colours are measured in a 
usefully consistent way, e.g.\ at the same position and with the same
profile model, across all available passbands. In many ways, 
entities \verb+ListRemeasurement+ and \verb+SourceRemeasurement+
are analogous to \verb+Detection+ and \verb+Source+ respectively and
hence show similar relationships between each other and 
\verb+MultiframeDetector+. However, every \verb+SourceRemeasurement+
is driven by one \verb+Source+ -- this defines the one--to--one
relationship between these entities. Futhermore, certain photometric
attributes of the remeasurement entities will have slightly 
different meanings to their analogues in \verb+Detection+
and \verb+Source+, most notably flux measurements at positions
defined by the driving list. In order to cope with the possibility
of marginally detected or negative fluxes, one approach (which has
yet to gain wide acceptance in the astronomical literature)
is to adopt the magnitude scale of Lupton, Gunn \& Szalay~(1999) in the
remeasurement entities for any calibrated flux attributes to be
usefully defined in such a situation. (We note that at the time of
writing, list--driven photometry has yet to be implemented in the
WSA).

It is important to note that the WSA is required to track a number
of different science programmes in which the prescription for
source merging (i.e.\ the required filters and number of distinct
epoch passes in those filters) will be different. Before illustrating
a specific example of the application of the generalised catalogue
ERM, we need to discuss the top--level data model of the WSA that
describes the observational programmes contained within it.

\subsubsection{Top--level metadata}
\label{topLevel}

\begin{figure}
\includegraphics[width=70mm,angle=0]{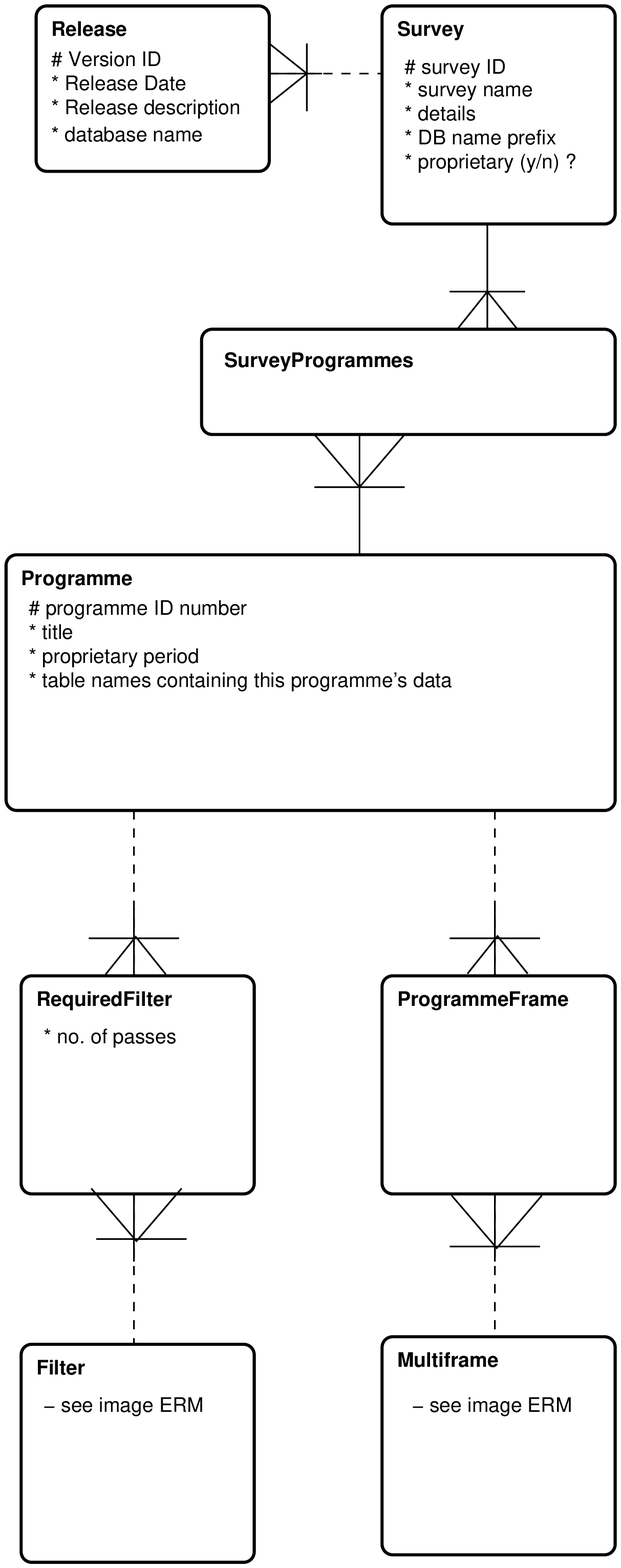}
\caption{Relational model for WFCAM surveys and
programmes in the WSA (see text).}
\label{surveyProgERM}
\end{figure}

\begin{figure}
\includegraphics[width=84mm,angle=0]{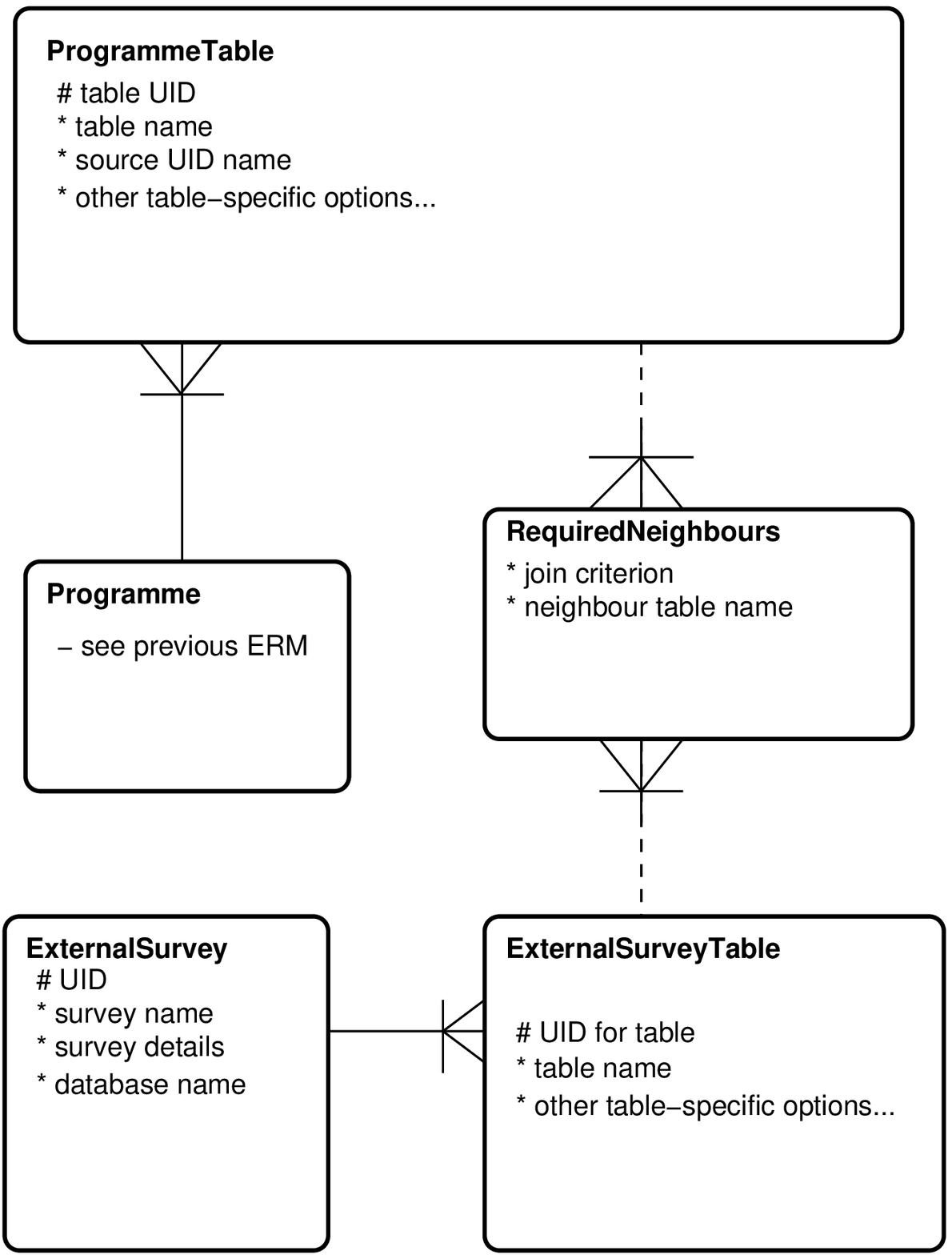}
\caption{Relational model for WFCAM and external survey 
catalogue metadata entities, and joins between them (see text).}
\label{joinERM}
\end{figure}

In order to track the various programmes for which the WSA is
required to hold data, e.g.\ survey (UKIDSS), non--survey (private
proprietary) and `service' programmes, the set of entities in the
schematics in Figures~\ref{surveyProgERM} and~\ref{joinERM} have
been identified. Consider the UKIDSS survey, which consists of
five sub--survey components. Once again, in simple relational
terms we identify entity \verb+Survey+ with a mandatory one--to--many
relationship to a set of \verb+Programme+s, e.g.\ the UKIDSS
LAS, GPS, GCS etc.,
with each \verb+Programme+ consisting of one or more
\verb+Multiframe+s. Note, however, in this case the relationships
between \verb+Survey+ and \verb+Programme+, and \verb+Programme+
and \verb+Multiframe+, are propagated via two further entities,
\verb+SurveyProgrammes+ and \verb+ProgrammeFrame+ where the latter
have optional or mandatory many--to--one relationships with their
linked entities. 
In the case of entity \verb+Programme+, the 
generalisation in its relationship to \verb+Multiframe+
allows each image dataset in the latter to belong to none, one,
or more than one \verb+Programme+. This is useful, for example,
in the UKIDSS GPS and GCS \verb+Programme+s which overlap in
their surveyed areas, filter coverage and depth, and it is
clearly advantageous to use the same data for both rather than
duplicate survey observations. Note also that entity
\verb+RequiredFilter+ in Figure~\ref{surveyProgERM} specifies
the prescription for source merging for a given \verb+Programme+,
where every \verb+Programme+ {\em may} have one or more
\verb+RequiredFilter+s specified. For example, the UKIDSS LAS requires
filter combination YJHK with two passes at J, whereas certain
non--survey \verb+Programme+s may not require source merging at all.
Every \verb+RequiredFilter+ must of course reference an existing
\verb+Filter+, hence the mandatory many--to--one relationship
between those two entities. Finally, entity \verb+Release+ tracks
information about releases that have occured for a given survey;
every \verb+Survey+ {\em may} have one or more releases.

Figure~\ref{joinERM} shows the other main aspects of the WSA top--level
data model with relevance to the end--user. The WSA holds local copies
of external datasets, from various sources, as specified by the UKIDSS
consortium early in the requirements capture phase of the project.
These large datasets were anticipated as being essential to certain
science applications of the infrared surveys, and include the Sloan
Digital Sky Survey catalogue data releases, e.g.~Data Releases 2, 3~and~5
(Abazajian et al.~2004; Abazajian et al.~2005; Adelman--McCarthy et al.~2007); 
the 2MASS point and extended source catalogues
(Skrutskie et al.~2006); 
and the SuperCOSMOS Science Archive database (e.g.\ Hambly
et al.~2004b). The data model in Figure~\ref{joinERM} illustrates
that every \verb+ExternalSurvey+ consists of one or more
\verb+ExternalSurveyTables+ (e.g.\ 2MASS contains distinct point
and extended source tables) and every \verb+Programme+ has one or
more \verb+ProgrammeTables+ that are required to be joined in
pairs as specified in \verb+RequiredNeighbours+ (the joining
philosophy and procedure is discussed further in 
Section~\ref{neighbs} and in detail in 
Section~\ref{neighbours}). For example, the science requirements
for the UKIDSS LAS specify that the LAS merged source list should
be joined to the corresponding list in the SDSS. The generalisation
using entities \verb+ProgrammeTable+ and \verb+ExternalSurveyTable+
allow for arbitrary joins between {\em any} tables in the linked
surveys rather than linking \verb+Programme+ and 
\verb+ExternalSurvey+ directly which would result in only one
join being allowed for each pair of \verb+Survey+s.

\subsubsection{Example data model for programme catalogue data}
\label{exampleCat}

\begin{figure*}
\includegraphics[width=120mm,angle=0]{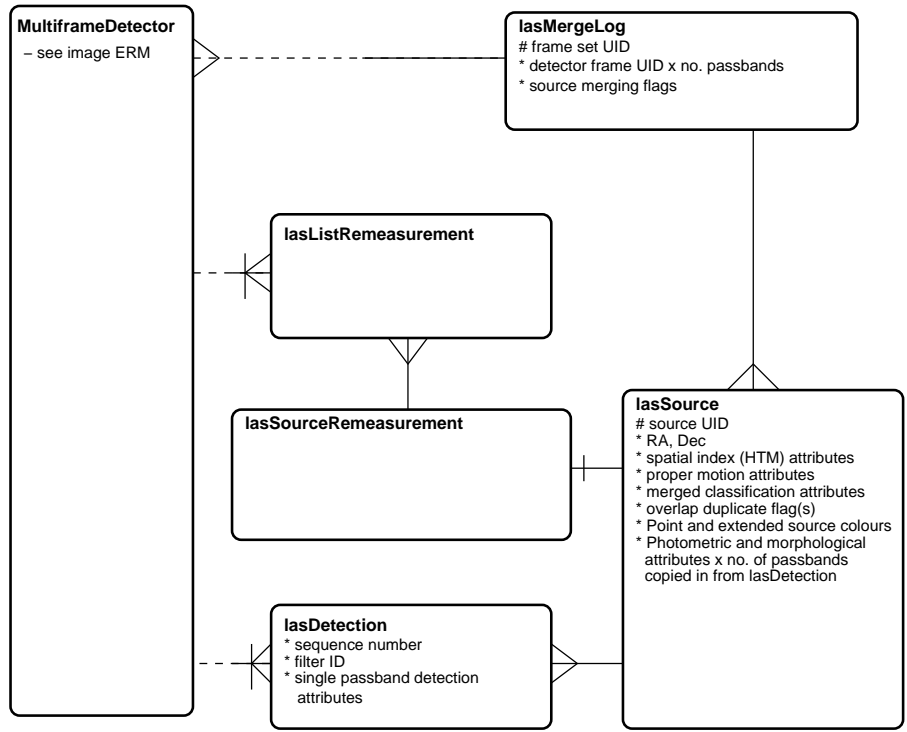}
\caption{Relational model for UKIDSS LAS catalogue
data in the WSA, following on from the general
case in Figure~\ref{catERM} and discussed in
Section~\ref{genericCat}.}
\label{lasERM}
\end{figure*}

The previous Section illustrates the hierarchy of \verb+Survey+s,
\verb+Programme+s and their associated descriptive data model.
It should be clear now that a distinct entity for each of
\verb+Source+, \verb+MergeLog+ and \verb+SourceRemeasurement+
(Figure~\ref{catERM}) is implied for every \verb+Programme+,
since the prescription for source merging in \verb+RequiredFilter+
will be different in each case and the attribute sets in these
three merged source entities will be different (imposing the same
attribute set on all merged source entities would necessitate a
large number of defaults, i.e.~unused attributes, for most). 
In fact each UKIDSS survey \verb+Programme+ tracked
in the WSA has the set of five entities shown in Figure~\ref{catERM}
for the purposes of storing catalogue data. This is because the single
passband entities \verb+Detection+ and \verb+ListRemeasurement+
are closely related to their respective merged source entities
within a given programme, and because it can aid performance
and housekeeping if large datasets are split into
related subsets (in addition to clarifying the data model
for the end user). In Figure~\ref{lasERM}
we give a specific example of the catalogue data model for the
UKIDSS LAS. (Note that non--survey \verb+Programme+s do not include
remeasurement and merged source entities unless these are
requested by their PIs). 
In addition to  the general description already given 
in Section~\ref{genericCat}, it is worth noting at this point
that we {\em denormalise} \verb+lasDetection+ and \verb+lasSource+
in that a small subset of the most useful single passband
photometric attributes are copied from the former into the latter
to facilitate simple end--user querying of what are anticipated
to be the main science tables for the survey datasets, in this
case the merged source table \verb+lasSource+. For more details
concerning source merging, see Section~\ref{cu7}

\subsubsection{Calibration data model}

Pipeline processing delivers {\em instrumental} astrometric
and photometric attributes and calibration coefficients
(Irwin et al.~2007). For example, each single passband
detection comes with an (x,y) coordinate location, and
a set of FITS World Coordinate System (WCS; 
Calabretta \& Griesen 2002) comes with each image for
transformation to celestial coordinates. Photometric
attributes are also supplied as instrumental fluxes along
with a set of calibration coefficients for each image 
(zeropoints, aperture corrections, etc.) to be applied to
put the photometric quantities on a standard magnitude
scale. The WSA stores all this information, and stores
calibrated quantities according to the 
the current calibration in further attributes for ease of
use. Hence, entity \verb+Detection+ (Figure~\ref{catERM})
contains (x,y) and flux attributes along with 
(RA, Dec, $l$, $b$, $\lambda$,
$\eta$)\footnote{($\lambda$, $\eta$) are spherical polar
survey co-ordinates defined for the SDSS}
celestial coordinates 
and a calibrated magnitude for every flux (and flux error)
attribute.

\begin{figure}
\centerline{\includegraphics[width=45mm,angle=0]{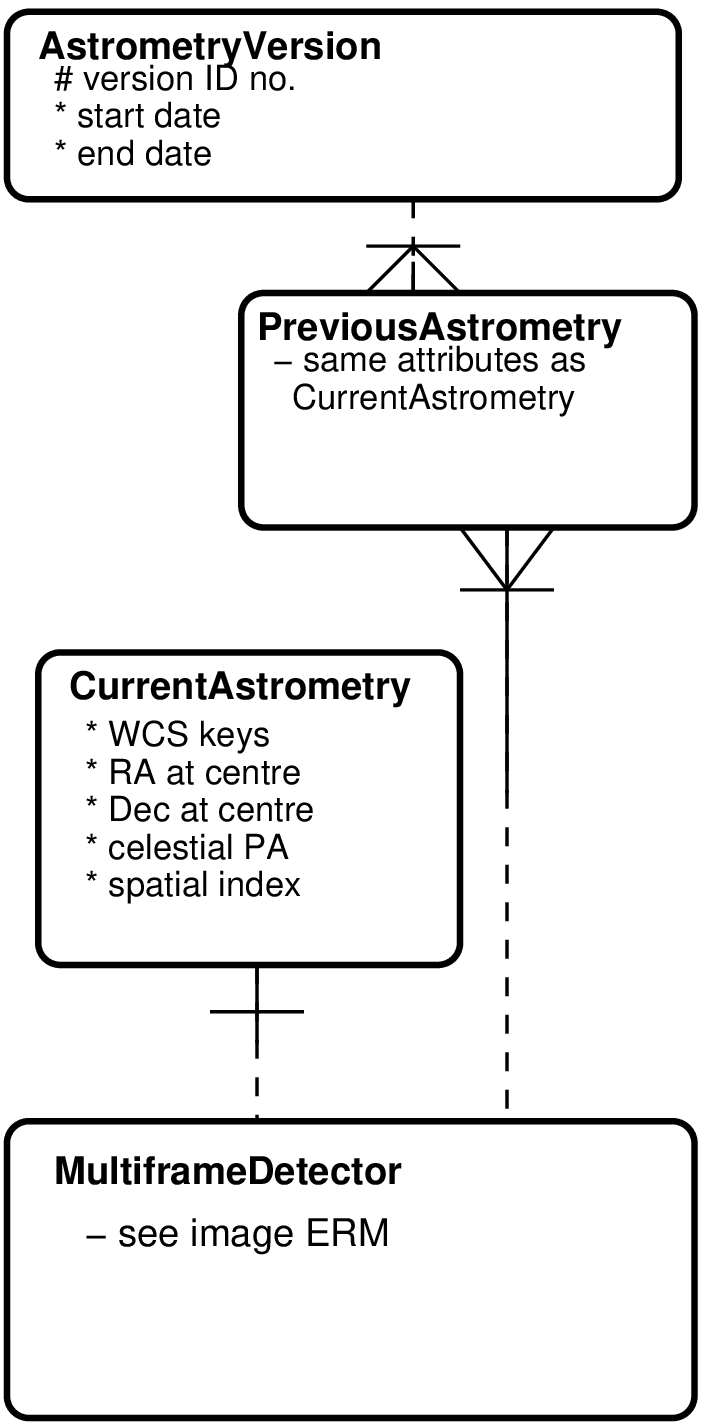}}
\caption{Relational model for astrometric calibration data
in the WSA. Entities and attributes are included to allow
for recalibration.}
\label{calERM}
\end{figure}

The advantage of storing instrumental quantities and calibration
coefficients is that updates to the calibration can be
tracked -- e.g.~at some point in the future, when a greater
understanding of the WFCAM instrumental behaviour has been
gained and a much larger amount of data is available, it may
be possible to recalibrate astrometry and photometry. Moreover,
for photometry in particular, additional calibration constraints
(e.g.~over many nights, or employing overlap regions between
adjacent frames) are available within the WSA that are not 
easily implemented in nightly pipeline processing. In 
Figure~\ref{calERM} we show the relational model for astrometric
calibration data to illustrate the approach (photometric
coefficient attributes are contained within the entities
\verb+Multiframe+ and \verb+MultiframeDetector+ already
identified in Figure~\ref{imageERM}). Astrometric calibration
coefficients are stored in entity \verb+CurrentAstrometry+ which
has a one--to--one relationship with \verb+MultiframeDetector+,
optional on the side of the latter. These coefficients, and some 
attributes calibrated using them, are gathered together in this
entity to make recalibration more efficient; the optional
relationship with \verb+Multiframe+ reflects the fact that not
all frames are necessarily astrometric (e.g.~darks). The other
two entities are included to track recalibration (if/when that
occurs): each \verb+MultiframeDetector+ {\em may} have one or
more \verb+PreviousAstrometry+ calibrations; and each of the
latter must be identified with an \verb+AstrometryVersion+.
These last two entities are unlikely to be of use to the end
user but are included to illustrate the recalibration aspect
of the WSA functionality. Similarly, instrumental photometric
calibration attributes are unlikely to be used in most
end--user usage modes.

\subsubsection{Data model for neighbouring sources from catalogue joins}
\label{neighbs}

\begin{figure}
\includegraphics[width=84mm,angle=0]{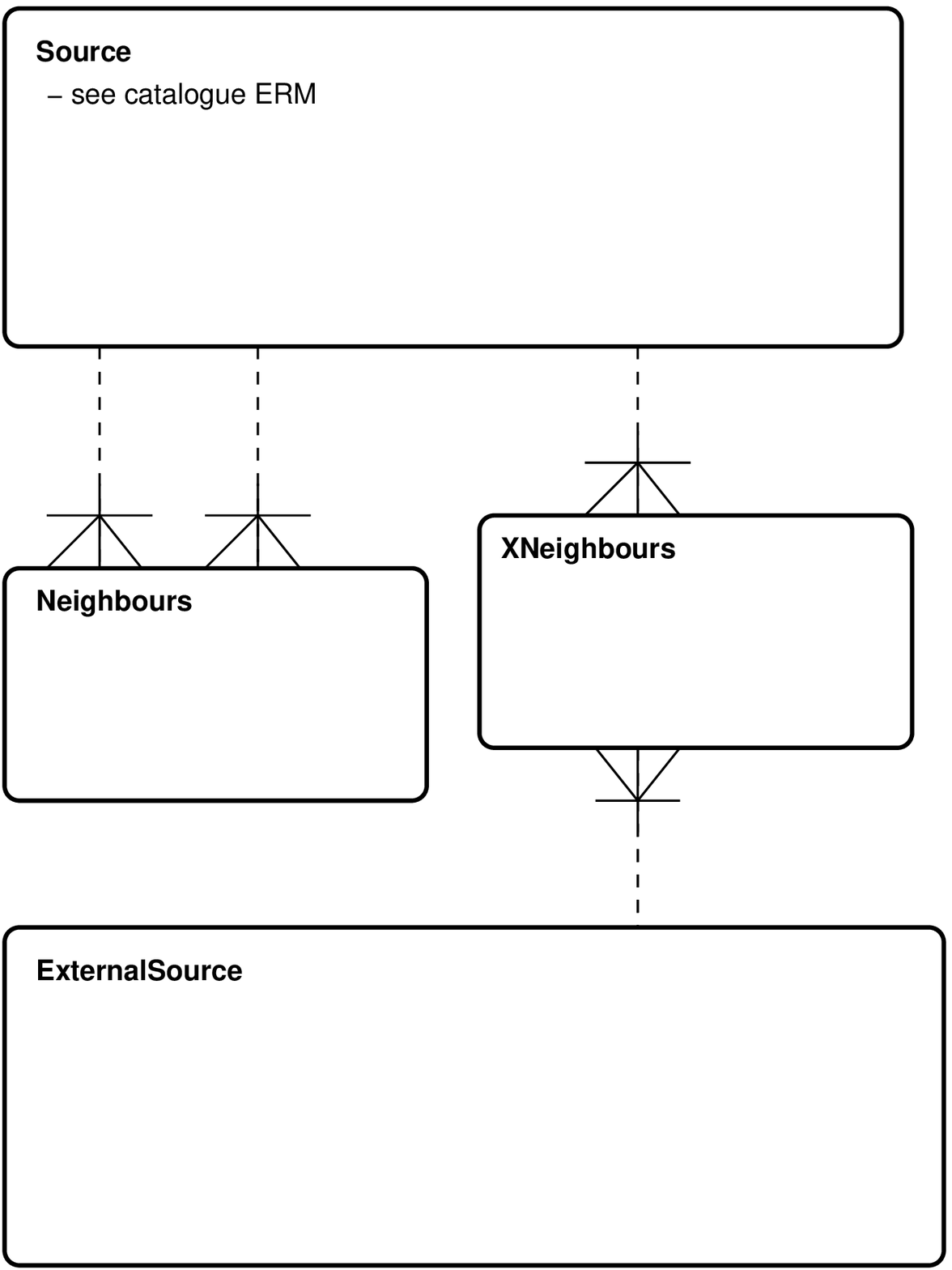}
\caption{Relational model for neighbouring sources
within a WFCAM source table, and between that
table and an externally derived source list
(e.g.~an optical survey).}
\label{neighbsERM}
\end{figure}

As already indicated in Figure~\ref{joinERM} and Section~\ref{topLevel},
the WSA is required to hold local copies of large survey datasets
produced elsewhere to facilitate cross--matched usage modes within the
archive system. In the general case, we ideally want some method of
associating all {\em nearby} sources between two lists rather than
merging the lists with some specific procedure that uses, for
example, positional coincidence within a small, fixed tolerance to
make one association for what is assumed to be the same object in
each. Positional errors are non--linearly dependent on brightness;
stellar positions change with time due to proper motion; some
usage modes may require {\em nearby} sources, as opposed to the
nearest or coincident source in two datasets. For these reasons,
the WSA follows the SDSS system of defining {\em neighbour}
tables when joining any two datasets where the 
scientifically useful neighbourhood around
any given object is defined by a maximum 
angular radius. The generalised relational
model of neighbour entities is shown in Figure~\ref{neighbsERM}.
Every WFCAM \verb+Source+ {\em may} have one or more 
cross--neighbours recorded in \verb+XNeighbours+ (one entity
for each cross--correlated \verb+ExternalSource+ is required).
An analogous relationship exists between the cross--neighbour
entity and the external source entity, i.e.~a many--to--one
relationship, optional on the side of \verb+ExternalSource+,
since once again every externally catalogued source {\em may}
be a neighbour of one or more WFCAM catalogued objects in
\verb+Source+.

Figure~\ref{neighbsERM} also models entity \verb+Neighbours+ which
is related to \verb+Source+ only. This is a {\em neighbour} table:
it is analogous to entity \verb+XNeighbours+, but it records
neighbours within \verb+Source+ for every object recorded in the
same entity. Hence, two optional one--to--many relationships exist
between \verb+Source+ and \verb+Neighbours+ since every 
\verb+Source+ {\em may} have one or more \verb+Neighbours+ while
at the same time every \verb+Source+ {\em may} be a neighbour of
one or more other \verb+Source+s. The concept of neighbour tables
is discussed in more detail in Section~\ref{neighbours} with
specific examples, and usage modes are illustrated in 
Section~\ref{queries}.

\subsubsection{Synoptic survey data model}
\label{synoptic}

\begin{figure}
\includegraphics[width=84mm,angle=0]{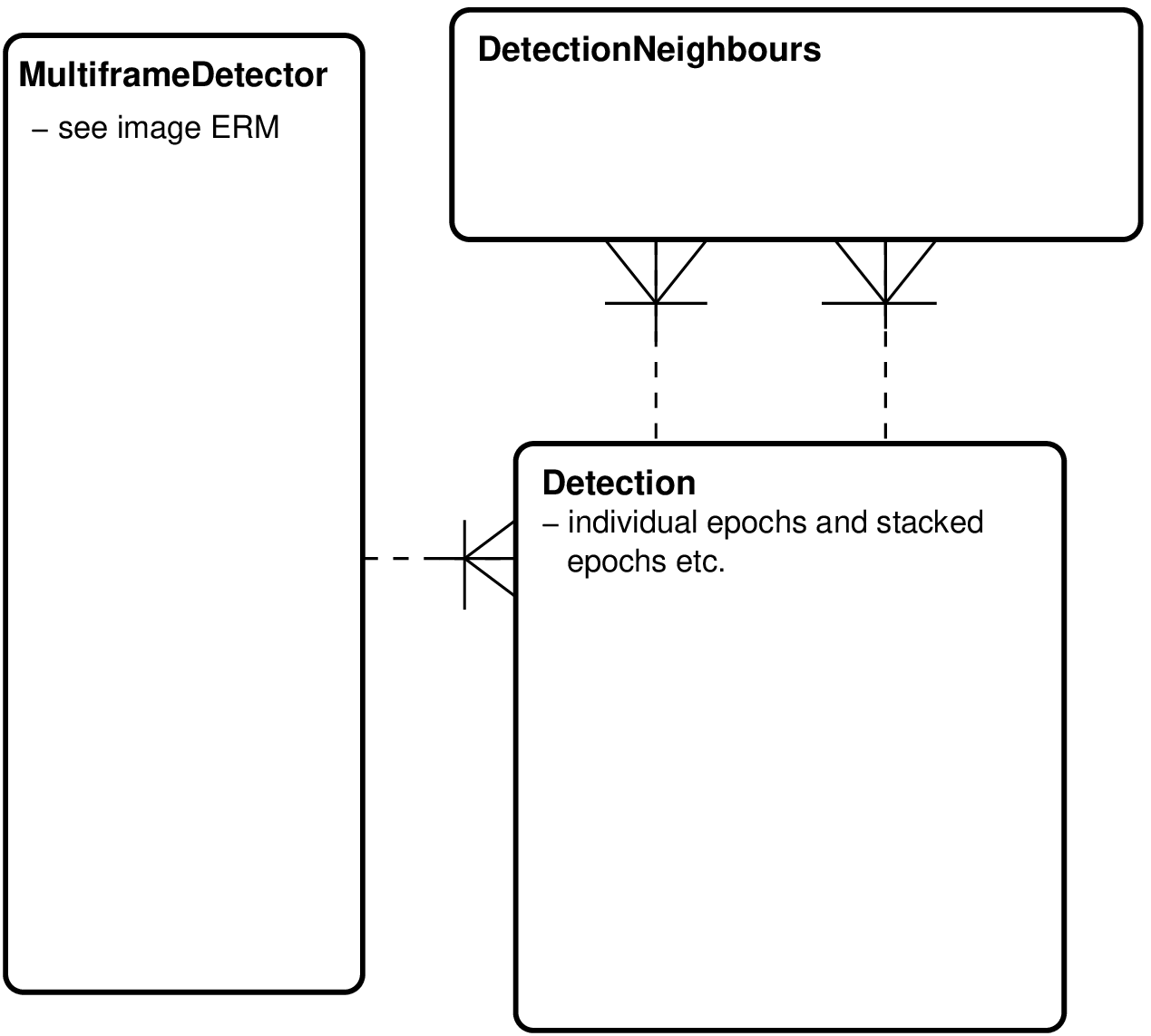}
\caption{Relational model for single--passband synoptic survey
data in the WSA, discussed in Section~\ref{synoptic}.}
\label{synopticERM}
\end{figure}

In Figure~\ref{catERM} we illustrate a data model that includes
provision for a merged source catalogue having a small, fixed number
of passbands/epoch visits via entity \verb+Source+. For example, the
UKIDSS LAS \verb+Source+ prescription is for visits in~YJHK with a
second epoch in~J. Modern imaging surveys, however, increasingly
aspire to extensive sampling of the time domain (e.g. Pan--STARRS,
Kaiser~2004; GAIA, Perryman~2005; LSST, Claver~2004), and we note that
both WFCAM and VISTA synoptic infrared surveys are being undertaken.
Such surveys, which have an indefinite and large number of field
revisits, require modifications to the data model presented in
Figure~\ref{catERM}. Figure~\ref{synopticERM} shows a single--passband
synoptic survey data model, where we have imaging
\verb+MultiframeDetector+s giving rise to one or more \verb+Detection+s
as before.
%but in this case the merged entity \verb+Source+ and its
%one--to--many relationship with \verb+Detection+ have a different
%purpose. Entity \verb+Source+ contains a master source list (perhaps
%generated from a stack of a subset of the available frames in each
%field) for each field revisited, and the one--to--many relationship
%with \verb+Detection+ tracks each individual epoch/passband measurement
%for each object. Attributes in \verb+Source+ include mean properties
%averaged over all available epochs (colours, classifications etc.),
%summary statistics (e.g.\ statistical variability indicators) and model
%fit quantities (e.g.\ position, proper motion and parallax from astrometric
%solutions). 
%Entity \verb+SourceEpochs+ provides storage for
%time--resolved photometric properties on a given timescale, for example
%colours at a given epoch -- i.e.\ those attributes that do not fit
%naturally into \verb+Detection+ or \verb+Source+. 
%Finally, two
%{\em neighbour entities} (Section~\ref{neighbs}) are defined in the
%data model. In the case of cross--neighbour entity 
%\verb+SourceXDetection+, provision for associations between the master
%source list and the individual epoch/passband measurements is made
%in order to facilitate simplified querying: for a given master source,
%the cross--neighbour list links to all available detections over all
%epochs and passbands in each case. Furthermore, neighbour entity
Neighbour entity
\verb+DetectionNeighbours+ provides links between each detection and all
other detections of that same object
in each case. 
%This may seem redundant given the master
%source and cross--neighbour entities, but it allows, for example, for
%very brief optical transients or very high proper motion stars that
%may not appear in the master source list, depending on exactly how the
%latter is defined. 

The basic relational design for synoptic survey data illustrated
here is appropriate
for a single--passband transit survey. However, it has a number
of disadvantages, including a large level of repeated associations
in the \verb+DetectionNeighbours+ entity. For~$N$ visits in a
given field, there will be at least $N\times(N-1)$ rows in
the neighbour table for every source since every combination
of the $N$ detections taken two at a time is listed. Moreover, if
the survey design is multicolour in~$M$ passbands,
\verb+DetectionNeighbours+ would rapidly becomes unmanageable as
every combination of $N\times M$ taken two at a time is recorded,
yielding $NM(NM-1)\approx N^2M^2$ entries for every source. This 
is addressed in the revised data model for VISTA synoptic surveys
presented in Cross et al.~(2007).

\section{Implementation}
\label{browser}

The relational data models presented previously are amenable to 
implementation in any RDBMS. The WSA is deployed in a commercial
software product, Microsoft `SQL~Server', a
system that is suitable for medium to
large--scale applications (this choice was made not least because
the SDSS Sky Server catalogue access systems are deployed on the
same -- see Thakar et al.~2003b). The implementation of the ERMs yields 
a set of database `objects' known as a {\em schema\/}. The database
objects mainly consist of tables, where each entity identified
previously maps to a table in the schema. These tables hold
the astronomical information (amongst other data) and can be
queried via the WSA user interface applications.

The WSA 
provides\footnote{\tt http://surveys.roe.ac.uk/wsa/www/wsa\_browser.html}
a {\em schema browser} which gives extensive information on the objects
(most notably the tables) in all available databases. The
schema browser initially presents the user with a tree--view
of databases that are held in the archive. Expanding any one
database item yields a sub--tree of objects (also expandable)
that includes the items described below.

\subsection{Tables and indexes}

These browser entries are the primary source of astronomical
information for users. Table names are self--explanatory and
indicative of their associated data model entities presented
previously (e.g.~\verb+dxsSource+, \verb+gcsSource+,
\verb+lasSource+ hold merged multi--colour source entries
as modelled in Figure~\ref{catERM} for the UKIDSS DXS, GCS
and LAS respectively). Clicking on any table name yields a
full description of the table and its columns, including
attribute names, data types, units and default values.
Further information is available for some attributes (those
having small icons) that link to brief `tool--tip' style
pop--up windows and glossary entries that provide more
detailed information (e.g.~for standard pipeline processing
catalogue attributes, a summary of relevant algorithmic
details is available -- see, for example, those for
\verb+gauSig+, \verb+aperFlux1+ and \verb+class+ etc.\
in \verb+lasDetection+). Finally, a small but nonetheless
important detail is that some attributes in a table's list of
columns have highlighted background colours in the browser.
This indicates that an {\em index} exists in the RDBMS for
that attribute: execution of queries predicated on
indexed quantities is very efficient.

\subsection{Views}

Views are simply definitions of tabular sets of data derived from
the tables available in the database, and can be queried in the
same way as those tables. A view may be a subset of a single 
table (i.e.\ a subsample of the rows and/or columns available)
or a superset of several tables. Views enhance the schema over
and above the set of tables without incurring any storage 
penalty in the RDBMS system since the underlying tables are
accessed at query time for the defined view row/column set. 
As far as the user is concerned, a view is simply a convenient
way of accessing, via a single short name, a set of data formed
from a selection made from one or more other database objects
(normally tables). In the WSA schema browser, expanding the
view tree of a given database produces the list of available
views defined within it; clicking on a given item produces a
description and the formal (SQL) definition of the view.
Examples of views in recent UKIDSS database releases are:
\begin{itemize}
\item \verb+lasPointSource+ -- a subsample of \verb+lasSource+
rows containing point--like sources in the UKIDSS LAS;
\item \verb+lasYJHKMergeLog+ -- a subsample of \verb+lasMergeLog+
rows containing frame sets with complete YJHK filter coverage in
the UKIDSS LAS;
\item \verb+lasYJHKSource+ -- a subsample of \verb+lasSource+
rows containing objects in areas with complete YJHK filter
coverage in the LAS.
\end{itemize}
Other views are defined for the UKIDSS databases, for example
views that select samples trading off completeness versus
reliability -- consult the schema browser for more details.
The view definitions also serve as examples illustrating
the SQL syntax required to make a specified selection
(but more of this later).

\subsection{Functions}

Some useful astronomical functions are provided in certain WSA
databases, and these are listed in the browser tree--view under
`Functions' where available. Functions generally take as
arguments an attribute name list: for example, functions are
provided to convert RA and Dec expressed in decimal degrees
into a more conventional sexagesimal string. Other functions
include spherical astronomy routines (e.g.~computation of
great--circle distance between two points on the celestial
sphere) and utility functions to format standardised IAU
names for arbitrary sources based on equatorial spherical
co-ordinates. Once again, for more details see the schema
browser.

\subsection{Data manipulation: curation procedures}

The WSA design incorporates a set of curation application procedures
for the creation of science--ready database releases for users.
Curation procedures include transfer of pipeline--processed data,
ingest of those data into the DBMS, production of quick--look
images for browsing, and source merging. In this Section we give 
details of the most important procedures from the point of view of
the end--user. 

\subsubsection{Quality control}
\label{qc}

The design of the WSA includes provision of features to enable general
quality control (QC) of ingested data. Such features as a deprecation
code attribute in every table subject to ingest modification, and
expurgation of deprecated data in final released database products 
are provided. General QC is necessarily
a rather open--ended problem requiring much interaction with the data,
at least in the initial stages of survey operations. Although the
WSA design does not preclude fully automated QC procedures, presently the
UKIDSS data (for example) have a lengthy semi--automated QC process
applied, some details of which are given in Dye et al.~(2006) and
Warren et al.~(2007a). Table~\ref{qcCodes} provides details of the QC
checks applied to UKIDSS data as they stand at the time of writing.
Note, however, that for UKIDSS released database products all deprecated
data are removed, so users will see only those data records having
attribute \verb+deprecated+=0. Presently, none of the above QC procedures 
are applied to non--survey data held in the WSA.

\begin{table*}
\caption[]{WSA quality control deprecation codes and their meaning.}
\label{qcCodes}
\begin{tabular}{rl}
\hline
\multicolumn{1}{c}{Deprecation} & Description \\
\multicolumn{1}{c}{code}        &             \\
\hline
 1 & Stack frames that have no catalogue\\
 2 & Dead detector frames or all channels bad\\
 3 & Undefined and or nonsensical critical image metadata attributes\\
 4 & Poor sky subtraction (via pipeline sky subtraction scale factor)\\
 5 & Incorrect combination of exposure time/number/integrations for survey specific projects\\
 6 & Incorrect frame complements within groups/nights (for incomplete observing `blocks')\\
 7 & Undefined values of critical catalogue attributes for stacks\\
 8 & Seeing=0.0 for a stack\\
 9 & High value of sky that compromises the depth\\
10 & Seeing outside specified maximum\\
11 & Photometric zeropoint too bright\\
12 & Average stellar ellipticity too high\\
13 & Depth (as calculated from sky noise and 5$\sigma$ detection in fixed aperture)
    is too shallow compared to overall \\
 & histogram distribution (i.e. shallower 
    than ~0.5mag wrt the modal value) or sky noise is too high for sky level\\
14 & Default aperture correction outlying in distribution of same versus seeing\\
15 & Pipeline photometric zeropoint inconsistent between image, extension and/or catalogue extension keywords \\
16 & Difference in detector sky level wrt to mean of all 4 detectors is outlying
    in the distribution of the same\\
18 & Provenance indicates that a constituent frame of a combined frame product
    includes a deprecated frame\\
19 & Inconsistent provenance for a stack or interleaved frame indicating something
    wrong with the image product (usually \\
 & corrupted FITS keywords confusing the pipeline)\\
20 & Detector number counts indicate some problem, e.g.~many spurious detections\\
21 & 5$\sigma$ depth of detector frame more than 0.4mag brighter than modal value
    for a given filter/project/exposure time\\
22 & Astrometry check (pixel size and/or aspect ratio) indicates something
    is wrong with the image\\
26 & Deprecated because frame is flagged as ignored in pipeline processing\\
40 & Science (stack) frame is not part of a survey (e.g. high latitude sky 
    frames in the GPS)\\
60 & Eyeball check deprecation: trailed\\
61 & Eyeball check deprecation: multiple bad channels\\
62 & Eyeball check deprecation: Moon ghost\\
63 & Eyeball check deprecation: Sky subtraction problem\\
64 & Eyeball check deprecation: Disaster (catchall category for the indescribable)\\
65 & Eyeball check deprecation: Empty detector frame\\
66 & Flat fielding problem\\
70 & Eyeball check requires deprecation, but this is the best that can be done so this
    should not be reobserved \\
 & (e.g.\ very bright star in WFCAM field of view)\\
80 & Deprecated because observation (block, object, filter) has been repeated later
    (shallow surveys only). The latest \\
 & duplication in each case is kept\\
99 & Manually deprecated because of some data flow system issue (e.g.\ pipeline malfunction)\\
100 & Multiframe deprecated because all detectors have been previously deprecated\\
101 & MultiframeDetector deprecated because parent Multiframe is deprecated\\
102 & Detection deprecated because parent Multiframe Detector deprecated\\
$>127$ & Deprecated because pipeline reprocessing supersedes it (where value $=128+$ deprecation code as defined above)\\
255 & Deprecated database--driven product (e.g.\ deep stack)\\
\hline
\end{tabular}
\end{table*}

Furthermore, the WSA includes provision for quality bit flagging of
catalogue records in common with error condition flagging in 
similar survey projects and source extraction pipelines, e.g.~SDSS
(Stoughton et al.~2002), S--Extractor (Bertin \& Arnouts~1996) and
SuperCOSMOS Sky Survey source extraction (e.g.~Hambly et al.~2001b).
This procedure consists of the assignment of single bits to
represent Boolean true/false conditions in an integer attribute
modified during source extraction and/or post--processing of the
extracted catalogues. The WSA data model includes provision for
both, and Table~\ref{qbits} gives details of the post--processing
quality error bit flags currently defined. Following Hambly et 
al.~(2001b) and references therein, the philosophy is to use more
significant bits in the flag for more severe quality error
conditions. Hence the numerical value of the quality flag can be
used as a measure of the relative quality of that catalogue
record: the higher the quality error value, the more likely it
is that the record is spurious. Of course, individual quality bits
can be tested also to see if a given condition is true for a
catalogue record -- this is achieved using the appropriate bit
mask (expressed in hexadecimal in Table~\ref{qbits}).

\begin{table*}
\caption[]{Post--processing error quality bit flags currently assigned
in the WSA curation procedure for survey data. From least to
most significant byte in the 4--byte integer attribute 
({\tt ppErrBits}; see later), byte~0 (bits~0 to~7) corresponds to
information on generally innocuous conditions that are
nonetheless potentially significant as regards the integrity of
that detection; byte~1 (bits~8 to~15) 
corresponds to warnings; byte~2 (bits~16 to~23) corresponds to
important warnings; and finally byte~3 (bits~24 to~31) corresponds
to severe warnings. In this way, the higher the error quality bit 
flag value, the more likely it is that the detection is spurious.
The decimal threshold (column~4) gives the {\em minimum} value
of the quality flag for a detection having the given condition
(since other bits in the flag may be set also). The corresponding
hexadecimal value, where each digit corresponds to~4 bits in the
flag, can be easier to compute when writing SQL queries to test
for a given condition (see later).}
\label{qbits}
\begin{tabular}{crlrc}
\hline
Byte & \multicolumn{1}{c}{Bit} & Detection quality issue & 
       \multicolumn{1}{c}{Decimal}   & Hexadecimal \\
     &     &               & 
       \multicolumn{1}{c}{threshold} & bit mask    \\
\hline
0 &  0 & Close to a dither edge (not yet implemented) &  1 & 0x00000001 \\
0 &  2 & Near to a bright star (not yet implemented) &        4 & 0x00000004 \\
0 &  4 & Deblended                        &       16 & 0x00000010 \\
0 &  6 & Bad pixel(s) in default aperture &       64 & 0x00000040 \\
2 & 16 & Close to saturated               &    65536 & 0x00010000 \\
2 & 19 & Possible crosstalk 
          artefact/contamination          &   524288 & 0x00080000 \\
2 & 22 & Within dither offset of image boundary &  4194304 & 0x00400000 \\
\hline
\end{tabular}
\end{table*}

\subsubsection{Source merging}
\label{cu7}

Combining single passband and/or single epoch detections into a merged
multi--colour, multi--epoch record is one of the major curation activities
applied after ingestion of pipeline processed catalogues. The merging 
philosophy is based on a number of fundamental assumptions that are made
in order to provide a procedure that is scalable to {\em billions} of
individual object records. Primarily, source merging is based on the
concept of {\em frame sets} (e.g.\ Sections~\ref{genericCat} 
and~\ref{exampleCat}) where the individual passband/epoch detections to be 
merged are assumed to come from a set of well aligned frames. This has
the major advantage that given any one detection, the corresponding
detection in another filter or at another epoch is easily and quickly
locatable in a tiny subset of all available detections over all frames
since the procedure is restricted in its search to one specific frame.
One of the disadvantages is that if a survey area is tiled differently
between the various passband and epoch visits made, then this
assumption is invalid and unmerged detections will appear in the final
source list. Another less critical assumption is that a small subset
of individual passband/epoch detection attributes is propagated 
into the source table for
each merged source. This subset includes what is considered to be the
most useful subset of photometric, astrometric and morphological
attributes along with associated errors,
and currently includes a selection of four fixed aperture
and Petrosian flux measures, 
model profile flux estimators, individual passband/epoch
morphological classifications and image quality attributes. Note,
however, that all detection attributes are always available in the 
detection tables; propagating a few of those more commonly used 
simply makes end--user querying easier and faster.

In addition to propagating individual detection attributes, the source
merging procedure computes new attributes. For example, default point
and extended source colours and associated errors are calculated, in
pair combinations of filters adjacent in wavelength (e.g.\ for the
UKIDSS LAS YJHK data, colours Y--J, J--H and H--K are computed). Also,
a normally distributed merged classification statistic and
associated discrete classification code are calculated using the
available individual passband/epoch values. The standard
80--parameter detection attributes in the catalogue extraction software
(Irwin et al.~2007) include a normally distributed, zero mean, unit
variance statistic derived from the radial profile of each detected
object. This N(0,1) statistic describes how point--like each object
is with respect to an empirically--derived, idealised radial profile
set representing the PSF for the frame. A value of 0.0 indicates
ideally point--like, increasingly negative values indicate sharper
images (e.g.~noise--like), and increasingly positive values indicate
extended (e.g.~resolved galaxies). Because the statistic is normalised
over the full magnitude range of the data to the N(0,1) form, a 
selection between $\pm2.0$, regardless of magnitude, will yield a
sample notionally complete to 95\% for example. For merged sources,
a merged classification statistic is computed amongst those available
from the individual passband detections. This is computed as the sum
of those available, $n$, divided by~$\surd n$, noting that 
the result of averaging $n$ individual zero mean, unit
variance -- i.e.~N(0,1) -- statistics results in a distribution of 
RMS $1/\surd n$; hence rescaling the average by $\surd n$ -- or, 
equivalently, dividing the sum by $\surd n$ -- results in a combined 
statistic that is also N(0,1). Where a given passband
and/or detection is unavailable, or where calculation of merged
attributes is not possible, default values (Section~\ref{notNull})
are used to populate the fields of records affected. A complete
description of the attributes in each merged source list is available
online at the WSA via the {\em schema browser} (see Section~\ref{browser}).

\begin{figure*}
\includegraphics[width=150mm,angle=0]{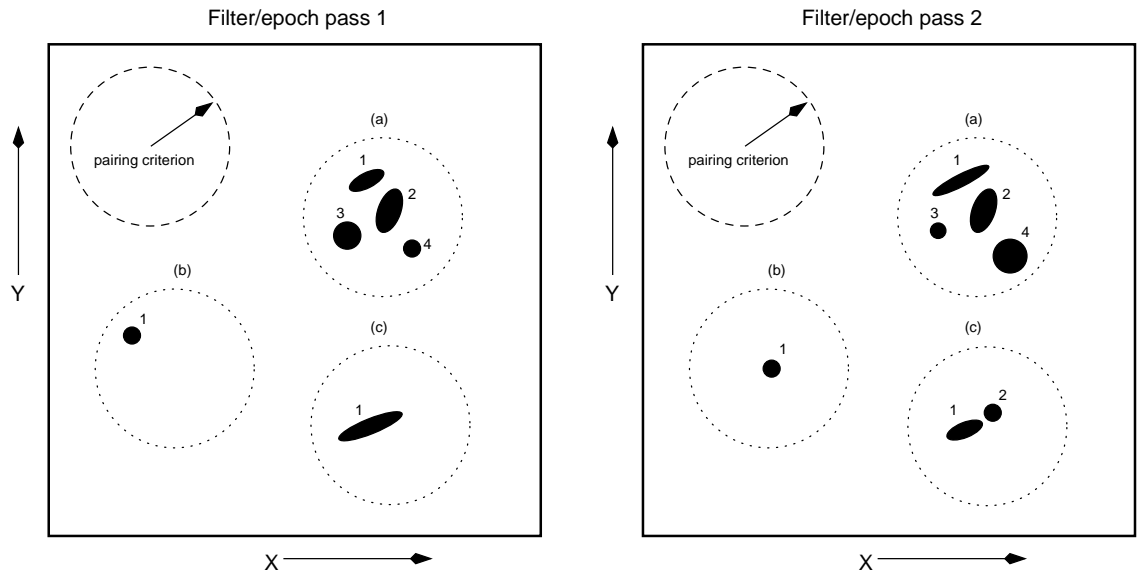}
\caption{Pretend catalogue data illustrating the core
pairing algorithm between two filter/epoch pass sets in a small
area: (a) close, but well separated objects paired 1a1$\equiv$2a1,
1a2$\equiv$2a2 etc.; (b) isolated moving object 1b1$\equiv$2b1;
(c) differently deblended objects, where 1c1$\equiv$2c1, 2c2 remains
unpaired since although 1c1 is within the pairing tolerance of 2c2
when set~2 is master, when set~1 is master 2c1 is closest to 1c1
and hence 1c1$\equiv$2c2 fails at the hand--shaking stage (see
text for more details).}
\label{pairOK}
\end{figure*}

\begin{figure*}
\includegraphics[width=150mm,angle=0]{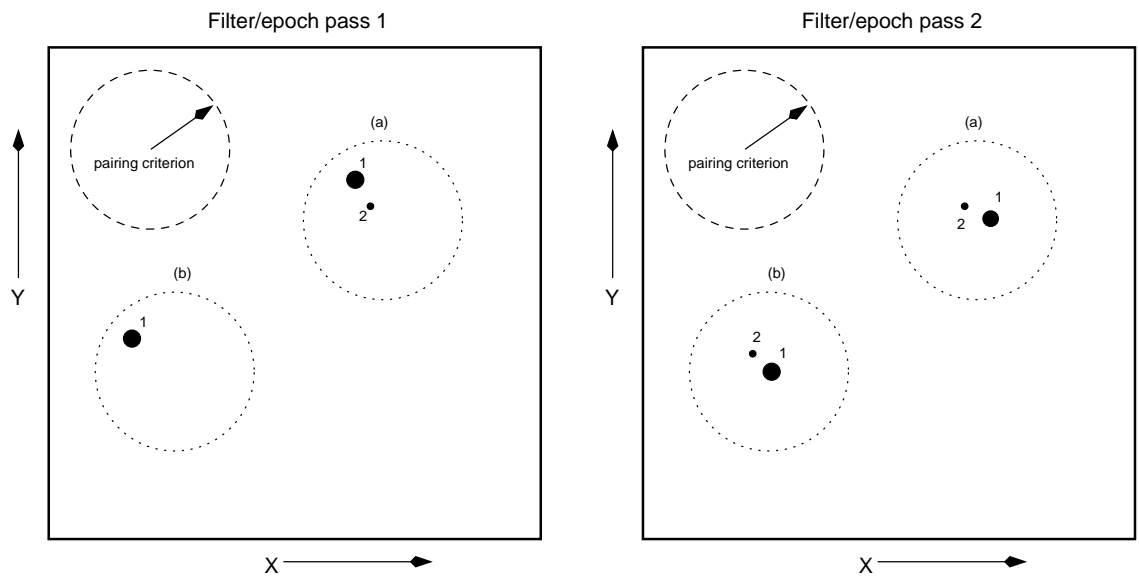}
\caption{As Figure~\ref{pairOK}, but illustrating some limitations of the
current WSA pairing algorithm: (a) very high proper motion star (1) moves
past another object detected in both sets -- 1a2$\equiv$2a2 satisfies
hand--shake pairing and is paired, but 1a1 points to 2a2 and 2a1 points
to 1a2 so the fast moving object is not correctly matched, fails
hand--shake pairing and remains unpaired; (b) very high proper motion object
moves past an object detected only in set~2 -- 1b1$\equiv$2b2 satisfies
hand--shake pairing and is incorrectly matched, while 2b1 remains
unpaired.}
\label{pairnotOK}
\end{figure*}

At the core of the WSA source merging procedure there is an efficient
pairing algorithm which associates detections between a given pair of
passbands/epochs based on proximity within a matching tolerance, or
pairing criterion. Table~\ref{pairtols} gives the 
radial pairing criteria currently
employed in UKIDSS source merging (these values are stored in the database
in table \verb+Programme+, attribute \verb+pairingCriterion+ for every
survey and non--survey programme that requires source merging).
Note that these tolerances are large
compared with the typical astrometric errors ($\sim0.1$\arcsec) to allow,
for example, for pairing of moving sources and very faint sources with
larger centroiding errors. Positional offset attributes for each 
filter/epoch pass are propagated into the merged source tables to
allow filtering of the merged source list at query time if a tighter 
pairing criterion is required (see later).
\begin{table}
\caption[]{WSA radial pairing tolerances used in UKIDSS source merging.}
\label{pairtols}
\centerline{\begin{tabular}{cc}
\hline
Survey & Radial pairing\\
       & criterion (arcsec)\\
\hline
LAS & 2.0 \\
GPS & 1.0 \\
GCS & 2.0 \\
DXS & 1.0 \\
UDS & 1.0 \\
\hline
\end{tabular}}
\end{table}
Once again, scalability becomes a major issue in a computationally
expensive procedure like record matching. The WSA philosophy
necessarily requires a compromise between speed and 100\% accurate
source association for real data (with all its vagaries) in every
conceivable situation. Figure~\ref{pairOK} illustrates the
straightforward scenario where two passes over the same area of
sky are source merged. In order to correctly identify the {\em nearest}
match in each case, the pairing procedure creates a set of pointers
from set~1 as master to set~2 as slave, and in reverse from
set~2 as master to set~1 as slave. Then, a `hand--shaking' run
through the two sets of pointers is used to associate only those
matches that agree on each other being the nearest match. This
forward/reverse pairing and handshaking between any two detection
sets from different passes helps to reduce spurious matches to
a minimum -- case~(c) in Figure~\ref{pairOK}; case~(a) in 
Figure~\ref{pairnotOK} -- at the same time requiring only two
passes through the datasets.

Of course, this approach has its limitations. In Figure~\ref{pairnotOK}
we illustrate a few relatively rare or pathological cases where the
pairing algorithm will fail. However, we note that in cases where
pairing fails, unpaired records will be propagated into the merged
source lists as single passband detections and the end--user always
has at their disposal the flexibility provided by the neighbour table
(Section~\ref{neighbours}) to associate unmatched records of the same
source using a more sophisticated algorithm that is appropriate
to the particular science application. Clearly it is better to minimise
spurious pairings with an efficient algorithm than to attempt to match
every last record correctly with an impractically time consuming
process and at the same time risking incorrect matches. In this respect,
the core pairing algorithm in the WSA is conservative.

Given a frame set of filter/epoch passes, source merging proceeds by
taking each combination in pairs (e.g.\ for a single epoch ZYJHK 
set, Z~would be hand--shake paired with Y, J, H and K; Y with J, H
and K; J with H and K; and finally H with K) in order to enable 
merging of sources even when they are detected in as few as any two
passes (note that epoch passes are treated in exactly the same way
as different filter passes).
Lastly, the full set of pointers is worked through, and
merged sources created using the pointer associations. Each detection
in each frame in the set is propagated once, and once only, into the
merged source list, either as part of a merged record or on its own
as a single passband detection. Offsets in local tangent plane
co-ordinates are stored in the merged source list; these quantify
the distance between the pairings, the shortest wavelength considered
as the reference position in each case. In the the single epoch ZYJHK
example above, handshake pairs between Z as reference and YJHK as
`slave' are propagated into the merged source list first, with offsets
from the Z position stored in attributes \verb+jXi+, \verb+jEta+,
\verb+hXi+, \verb+hEta+, etc. Then any remaining Y detections would be
considered as reference for JHK slaves, etc.

The combination of i) a relatively large radial pairing criterion,
ii) handshake pairing, and iii) storage of offset values between
pairs provides maximum flexibility for the end user. The large
pairing radius maximises the chances of moving objects or objects
with large centroiding errors being paired. At the same time,
the handshaking procedure minimises spurious pairings in ambiguous
situations and forces nearest neighbour matches to be chosen
always. Finally, the availability of the pairing offsets in the 
merged source list enables the end--user to `tune' the pairing
radius at query time -- limiting pairing offsets can be specified 
to a maximum allowed by the radial pairing tolerance, as
appropriate to the science application (see later).

Finally, the WSA merged source procedure has a `seaming' feature that
enables selection of a science--ready merged source sample. All
imaging surveys have some degree of overlap between adjacent fields,
perhaps by design (to enable cross--calibration for example) or
because of instrument design or guide star limitations. The WFCAM
focal plane array (Casali et al.~2007), consisting of $2\times2$
detectors spaced by $\sim95$\% of the detector width, automatically
produces overlap regions in survey areas tiled for contiguous
coverage. Moreover, at high Galactic latitudes in particular, guide
star limitations can result in overlap regions of increased size.
Because repeat measurements of the same objects provide scientifically
useful information, the WSA philosophy is to retain duplicates in the
merged source lists, noting by means of an attribute flag 
(see Section~\ref{queries}) when a particular source has duplicates
present, and if so, which measurement is considered to be the
`best'. A source is considered to be duplicated when an adjacent
frame set contains a source within 0.8\arcsec\ using the same
pairing/handshaking procedure described earlier.
Briefly, the decision logic behind the choice of the best
source examines each set of duplicates (there may be two or
more to choose between) on a source by source basis. Source records
having the most complete passband coverage are favoured primarily; 
when two or more source records all have the same number of passband
measures, the choice of primary source is based on position relative
to the edges of the corresponding image (detections furthest from
the edges are favoured) amongst the set of duplicates having the
fewest quality error bit flags set (Section~\ref{qc}).

\subsubsection{Enhanced image products}

Within UKIDSS,
the DXS and UDS include image data in the same pointing and same filter 
that are taken over many observing blocks on the same or different nights. 
Thus it is 
necessary to stack these data at the archive to produce final image products
of the required depth. Cataloguing of these deep image stacks is also 
performed at the archive end. In the case of the UDS, the cataloguing is 
performed on mosaics made up of the 4 pointings so that objects at the 
boundaries of each pointing are measured at the full depth of the survey and
are not broken up into pieces.

The DXS uses the same stacking and cataloguing code used in nightly
pipeline processing of
the shallow surveys (Irwin et al.~2007) but the UDS images have been
stacked and mosaiced by the UDS team (e.g.\ Foucaud et al.~2007) 
using the Terapix software SWARP (Bertin et al.~2002), 
and we have used Source Extractor 
(Bertin \& Arnouts~1996) to catalogue the UDS deep mosaics. Only
those intermediate stack images (i.e.~the stack products of individual
observing blocks) that pass standard survey quality control
(e.g.~Section~\ref{qc}) are included in deep stacks/mosaics in
the WSA.

\subsubsection{Neighbour/cross--neighbour catalogue joins}
\label{neighbours}

As described in Section~\ref{neighbs}, the concept of a 
{\em neighbour table} provides a generalised cross--matching facility that
can service diverse usage modes. The WSA philosophy is to provide
neighbour tables for each merged source table, in order to allow, for
example, easy and quick internal consistency checks on calibration.
Furthermore, cross--neighbour tables are provided between UKIDSS
source tables and a selection of other large external survey datasets,
again to facilitate rapid cross--matched astronomical usage modes.
%Table~\ref{extcats} gives details of the external surveys currently
%held locally in the WSA, along with the UKIDSS surveys joined and
%the neighbourhood radius used in each case. 
%\begin{table}
%\caption[]{External survey datasets held in the WSA, along with details
%of `neighbour' joins with the UKIDSS surveys.}
%\label{extcats}
%\begin{tabular}{ccccc}
%\hline
%External & Database & Reference & UKIDSS & Neighbourhood  & \\
%survey   &  name    &           & table  & radius (arcsec)& \\
%\hline
%SDSS--DR2 & BestDR2  & ?? & lasSource & 
%\hline
%\end{tabular}
%\end{table}
We note that the generic problem of cross--matching
very large datasets (i.e.\ those containing $\ge$~billions of rows) 
is receiving attention in the burgeoning Virtual Observatory
(e.g.~O'Mullane et al.~2005
and references therein); the WSA currently holds local copies of
user--required external datasets (e.g.\ SDSS catalogue data releases,
the 2MASS catalogues) in lieu of fast VO--implemented solutions. As
far as a scalable implementation is concerned, the WSA employs bulk
data egress/ingest facilities provided in the back--end RDBMS, and
an application making use of the `plane sweep' algorithm (Devereux
et al.~2005) for extremely fast cross--matching.

Further details concerning neighbour tables,
the external datasets held in the WSA and 
corresponding cross--neighbour tables are given online in the
{\em schema browser} (Section~\ref{browser}). For example, a
cross--matching neighbourhood radius of 10\arcsec\ is used generally
although this varies depending on the tables being matched.
Illustrative usage examples are given below.

\section{Illustrative science examples}
\label{queries}

Appendix~\ref{usages} lists some typical archive usage modes that were
identified in collaboration with the user community (i.e.\ the UKIDSS
consortium) early on in the WSA design phase. For casual browsing
and usage involving limited data subsets or very small areas of sky,
the static web forms provided in the WSA user 
interface\footnote{\tt http://surveys.roe.ac.uk/wsa/dbaccess.html}
are sufficient
to give the user the required data retrieval functionality. However,
for large--scale (e.g.~large area) and/or complex (e.g.~wholesale
statistical analysis) usage modes such as those illustrated in
Appendix~\ref{usages}, the provision to the user 
of a highly flexible interface is necessary. The WSA design
philosophy is to expose the Structured Query Language (SQL)
interface of the underlying RDBMS to the user to provide the 
required flexibility. 
%Insofar as SQL can be considered a simple programming language, a
Allowing users to execute data selections,
calculations and statistical computations on a machine
co--located with the data (i.e.~`server--side', or on the computer
that hosts the RDBMS itself) allows many users to access the large
data volume without recourse to wholesale distribution of the entire
data set.

A free--form SQL interface is 
provided\footnote{\tt http://surveys.roe.ac.uk:8080/wsa/SQL\_form.jsp}
in the WSA interface, and the example scripts below can be input
directly once a user is logged in and/or and appropriate database
release has been selected. Options within the interface include upload 
of a script file in addition to direct typing or cut--and--paste.
Note that the WSA free--form SQL interface imposes the following
limits on individual queries: maximum execution time 4800 seconds;
output rows$\times$columns~$=15\times10^6$ (i.e.~more attribute
columns selected implies fewer rows allowed in the results file).
These limits are imposed to prevent inexperienced users locking up
the service with erroneous and/or inefficient queries. No limit is
currently made on the number of concurrent queries or the frequency
with which they can be submitted.
Output formats include plain comma--separated text, FITS binary table
and VOTable\footnote{\tt http://www.ivoa.net/Documents/latest/VOT.html}, 
an XML format designed for international Virtual Observatory
(VO) initiatives. 

At the time of writing, other interface options are
under development; furthermore, the WSA is in the process of being
integrated into the VO via deployment of infrastructure developed by
the AstroGrid project (e.g.~Walton et al.~2006). In particular, 
UKIDSS database releases are published to the VO using
the AstroGrid Data Set Access (DSA) software. This has several advantages.
(i) The database is visible in VO resource registries around the world, and
so turns up in searches for databases of this kind. (ii) The metadata
describing the database (column names, unified content descriptors,
table structure) are available
through any VO-compatible software. (iii) Our database accepts queries in
the IVOA standard query format, Astronomical Data Query Language (ADQL).
This means that generic query software, such as the AstroGrid Query Builder,
can be used to issue queries to UKIDSS data. (iv) Our database understands
calls coming from libraries of routines in the ``Astro Runtime'', so that for
example, programmable use of the database can be made using high-level
languages such as Python.

\subsection{Guidance for the use of SQL in the WSA}

In Appendix~\ref{sql} we give a brief introduction to the 
fundamentals of SQL data retrieval (\verb+SELECT+) statements.
A more comprehensive guide is given 
online\footnote{\tt http://surveys.roe.ac.uk/wsa/sqlcookbook.html}
in the WSA `SQL cookbook', but in this Section we give brief
guidance to avoid common mistakes and to get the most from the
system.

\subsubsection{Use COUNT(*) and TOP N}

A good way of checking that a query is sensible is to replace the attribute
selection list with \verb+COUNT(*)+ since this skips creation of an output
file (including any DBMS look--up stage which can be time consuming for
large row counts) and can indicate if something is badly wrong in a
query (e.g.~an incorrectly specified table join). Consider query~B8
in Appendix~\ref{sql}, where a list of UKIDSS programmes/filters is
required:

\begin{verbatim}
SELECT COUNT(*)
FROM   Programme AS t1, RequiredFilters AS t2
/* NB: this is not a good query */
\end{verbatim}

\noindent
returns a count of~642 which is clearly wrong since there are five
UKIDSS programmes with on average~$\sim4$ filter coverage 
per programme -- we would expect a count of~$\sim20$. As noted
in Appendix~\ref{sql}, the related rows in the tables need to be
explicitly filtered using the referencing attribute common to
both -- in this case, the unique identifier \verb+programmeID+:

\begin{verbatim}
SELECT COUNT(*)
FROM   Programme AS t1, RequiredFilters AS t2
WHERE  t1.programmeID = t2.programmeID
\end{verbatim}

\noindent
returns a much more reasonable figure of~22 for UKIDSS DR2. Note that
summary counts for various survey release tables are available online
on the WSA web pages. Furthermore, data analysis plots showing the
density of stars and galaxies in colour space are also available --
these can be helpful when searching for rare objects in sparsely
populated colour ranges.

Note that another useful SQL command is \verb+TOP+ when debugging queries.
For example, \verb+SELECT+ \verb+TOP+ \verb+10 ...+ \verb+FROM ...+ 
will simply give the
first ten rows that satisfy the query and then execution will stop.
The reduced results set can be inspected for appropriateness and/or
errors before running the same query again without \verb+TOP 10+.

\subsubsection{Use GROUP BY for counts in arbitrary bins}

Following on from the use of \verb+COUNT(*)+, the addition of
\verb+GROUP BY+ (and furthermore statistical aggregates like 
\verb+AVG()+ for means, \verb+MIN()+ and \verb+MAX()+ for
minimum and maximum etc.~-- see Appendix~\ref{sql}) is very
useful for summarising the contents of a selection and/or
binning up data with a single pass through the table. For
example, what are the source counts in Galactic longitude
slices in the UKIDSS GPS? Do not use

\begin{verbatim}
SELECT COUNT(*)
FROM   gpsSource
WHERE  l BETWEEN 0.0 AND 1.0
\end{verbatim}

\noindent 
and then another query

\begin{verbatim}
WHERE  l BETWEEN 1.0 AND 2.0
\end{verbatim}

\noindent
and so on. It is much easier and much more efficient to use
\verb+GROUP BY+ to bin up in slices defined by longitude 
rounded to the nearest degree, for example:

\begin{verbatim}
SELECT CAST(ROUND(l,0) AS INT) AS longitude, 
       COUNT(*)
FROM   gpsSource
GROUP BY CAST(ROUND(l,0) AS INT)
ORDER BY CAST(ROUND(l,0) AS INT)
\end{verbatim}

\noindent
The query in Section~\ref{U8} below illustrates this further 
for the real survey data; for details of SQL functions
like \verb+CAST+ and \verb+ROUND+ consult the WSA online
documentation or any standard text on SQL.

\subsubsection{Take great care when joining tables}

Following on from checking using \verb+COUNT(*)+ as illustrated
above, in general follow these simple rules when employing
implicit table joins (i.e.~when supplying comma--separated lists
of tables in a \verb+FROM+ clause):
\begin{itemize}
\item for a list of~$N$ tables, ensure there are at least~$N-1$
\verb+WHERE+ conditions associating related rows in those tables;
\item never attempt spatial joins on co-ordinates (e.g.~the query
\verb+SELECT ...+ \verb+FROM+ \verb+lasSource+ \verb+AS s,+ 
\verb+lasDETECTION+ \verb+AS d+ with
an attempted joining clause of \verb+WHERE+ \verb+s.ra=d.ra+ 
\verb+AND+ \verb+s.dec=d.dec+
is inadvisable from many standpoints in addition to being dreadfully
inefficient);
\item always use the relational unique identifiers (i.e.~primary keys)
that associate related rows in related tables. 
\end{itemize}
For example, suppose a GPS user requires a source selection including
an attribute that is not available in the source table, e.g.~the modified
Julian date
of the~J observation and the isophotal magnitude in~H. The relational
model detailed previously shows that the related tables are
\verb+gpsSource+, \verb+gpsMergeLog+, \verb+gpsDetection+ and
\verb+Multiframe+, since every merged source belongs to a frame set
recorded in \verb+gpsMergeLog+ and consists of detections recorded
in \verb+gpsDetection+ arising from frames recorded in \verb+Multiframe+.
Examination of the arrangement of the UKIDSS data via the 
{\em schema browser} (Section~\ref{browser}) identifies tables
\verb+Multiframe+ and \verb+gpsDetection+ as containing the relevant
attributes \verb+mjdObs+ and \verb+isoMag+ respectively. Clearly,
these four tables must appear in the \verb+FROM+ clause of the
query and it is {\em vital} to include filters in the \verb+WHERE+
clause to associate the related rows:

\begin{verbatim}
SELECT TOP 10 s.sourceID, s.ra, s.dec,
       m.mjdObs AS jmjd, d.isoMag AS hIsoMag
FROM   gpsSource AS s, gpsMergeLog AS l,
       gpsDetection AS d, Multiframe AS m
WHERE  
/* Associate each source with its frame set: */
       s.frameSetID = l.frameSetID AND
/* Pick out the H band detection: */
       l.hmfID = d.multiframeID AND
       l.heNum = d.extNum AND
       s.hseqNum = d.seqNum AND
/* Pick out the J band frame: */
       l.jmfID = m.multiframeID AND
/* Keep only sources having J and H: */
       l.jmfID > 0 AND l.hmfID > 0
\end{verbatim}

\noindent
Note that the cross--referencing attributes are all defined as
{\em primary keys} in the referenced table entries in the
schema browser; the RDBMS is extremely efficient in locating rows
in tables using these.

\subsubsection{Use views, especially when new to the data}

There are a number of predefined selections based on various
optimisations of completeness versus reliability from cuts on
various morphological parameters available in survey database
releases in the WSA. Users are advised to check the available
views (again in the schema browser) when new to the WSA survey
datasets to see if there are any that suit a given astronomy
application. For example, there is a view of \verb+lasSource+
called \verb+reliableLasPointSource+, which is a predefined
selection with cuts on morphological parameters and a
requirement for detection in~Y, J~and~H for a reliable sample 
of point sources.

\subsubsection{Tune paired/cross--matched selections appropriately}

When using the merged source tables and/or the neighbour tables for
cross--matches between tables, users are advised to think carefully
about the maximum angular distance that is applicable to a given
astronomy application. The default pairing/cross--matching radii are
conservative in that they are set deliberately large to cover as
many applications as possible, but they may be too large for a
specific case and should be limited at query time. For example,
attributes \verb+Xi+ and \verb+Eta+ are available for each passband
in the merged source table -- if an astronomy application of the GCS
does not anticipate any pairings outside a 0.5\arcsec\ radius, then
the following predicates should be included:

\begin{verbatim}
WHERE zXi  BETWEEN -0.5 AND +0.5 AND
      zEta BETWEEN -0.5 AND +0.5 AND
      yXi  BETWEEN -0.5 ...
\end{verbatim}

\noindent
etc., for all passbands as necessary. For the case of cross--matched
selections employing neighbour tables, an appropriate limit on the
neighbourhood radius should be placed via a predicate on the
attribute \verb+distanceMins+ which is the distance in arcminutes
between any given `master' source and a `slave' cross--match in
the neighbourhood of the former. Further examples of this
are given below.

\subsection{Example SQL queries for astronomy usages}

In this Section we give 
%further introductory information specific to the WSA, followed by 
a set of astronomy SQL query
examples that are used as steps in part fulfilment of the
usages in Appendix~\ref{usages} where in each case, an explanation
is given and results are illustrated. As noted in Appendix~\ref{sql},
the WSA interface is case--insensitive: mixed case is used in the
examples for clarity in distinguishing SQL keywords and database
object names. Note also that \verb+/*...*/+ can be used to enclose
comments in the scripts; these are ignored by the WSA DBMS
when the script is run. The scripts are available 
online\footnote{\tt http://surveys.roe.ac.uk/wsa/pubs.html}
in the WSA documentation; further examples
of WSA SQL queries can be found in Dye et al.~(2006) and
Lodieu et al.~(2007a). The following queries are presented in order
of increasing complexity rather than in the order of the usages in
Appendix~\ref{usages}. Row counts and execution times at the end
of the scripts are those for UKIDSS Data Release 2 when selecting
FITS output format (for those queries that return many row
results sets).

\subsubsection{Candidate Galactic cluster members}
\label{U3}

\begin{figure*}
\centerline{\includegraphics[width=84mm,angle=270]{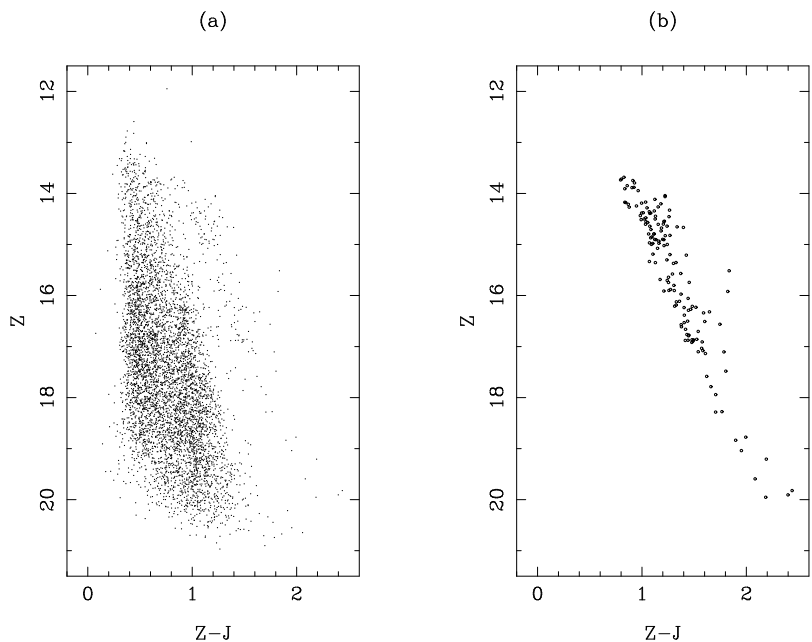}}
\caption{(a) Colour--magnitude plot of the results set from the
example query in Section~\ref{U3} without the final two predicates
showing the general field population and a clear brown dwarf
cluster sequence in the $\sigma$--Orionis cluster; (b) the same 
plot but using the two additional
predicates to select out the cluster sequence.}
\label{sigoplot}
\end{figure*}

Usage example~U3 in Appendix~\ref{usages} requires candidate cluster
member selection from the UKIDSS GCS by colour, magnitude and proper
motion. Colour selection is straightforward in SQL:

\begin{verbatim}
SELECT zAperMag3-jAperMag3 AS zmj,
       zAperMag3           AS z
FROM   gcsPointSource
WHERE
/* Positional cuts for the Sigma Orionis in the
   Orion Nebula Cluster (in degrees for both): */
       ra  BETWEEN +84.00 AND +85.00 AND
       dec BETWEEN  -2.85 AND  -2.30 AND
/* Magnitude cuts to avoid saturated sources: */
       zAperMag3   > 11.3 AND
       yAperMag3   > 11.5 AND
       jAperMag3   > 11.0 AND
       hAperMag3   > 11.3 AND
       k_1AperMag3 >  9.9 AND
/* Magnitude/colour cuts to select out the member 
   sequence: */
       zAperMag3 < 5.0*(zAperMag3-jAperMag3) + 
                                       10.0 AND
       jAperMag3-hAperMag3 > 0.3
/* UKIDSS DR2 rows returned: 144
   Execution time:           00m 12s  */
\end{verbatim}

\noindent 
where the colour/magnitude selection cuts have been defined by
examining colour--magnitude and colour--colour plots of the
selection made without the final two predicates. 
Figure~\ref{sigoplot} illustrates the results in~Z versus
Z$-$J colour--magnitude diagrams that clearly show the cluster
member sequence. At the time of writing, UKIDSS GCS proper motions
are unavailable because second epoch survey observations have yet
to start. However, Lodieu et al.~(2007a) show that, at least for 
brighter stars, proper motions can be computed by comparison with
2MASS catalogue positions; see also Lodieu et al.~(2007c) where
this kind of usage is deomstrated for the Pleiades open star
cluster in the UKIDSS GCS.

\subsubsection{Counts of objects that are unpaired between epochs}

Usage example~U4 in Appendix~\ref{usages} includes requirements to
select a sample of high proper motion stars having total proper
motion $\mu > 5\sigma_\mu$, and to count the number of sources
that are unpaired between  the two epochs of the UKIDSS LAS
J--band imaging. There are a number of ways of achieving this,
with increasingly sophisticated searches yielding increasingly
reliable candidates (but often at the expense of completeness).
As a first step, use of the view \verb+reliableLasPointSource+
is recommended. For the paired high proper motion selection,
we note that since
\begin{equation}
\mu^2 = \mu_\alpha^2 + \mu_\delta^2,
\end{equation}
where $\mu_\alpha$ and $\mu_\delta$ are the components of proper motion
(measured in the same units) in~RA and~Dec respectively, and combining
proper motion component errors in quadrature, we have that
\begin{equation}
\sigma_\mu = \frac{(\mu_\alpha^2\sigma_{\mu_\alpha}^2 +
                    \mu_\delta^2\sigma_{\mu_\delta}^2)^{1/2}}
                  {\mu}.
\end{equation}
Hence, the $5\sigma$ condition on total proper motion, $\mu > 5\sigma_\mu$
becomes
\begin{equation}
(\mu_\alpha^2 + \mu_\delta^2) > 5
 (\mu_\alpha^2\sigma_{\mu_\alpha}^2 + 
 \mu_\delta^2\sigma_{\mu_\delta}^2)^{1/2}.
\end{equation}

\noindent
In SQL, the high proper motion selection statement is

\begin{verbatim}
SELECT COUNT(*)
FROM   reliableLasPointSource
WHERE  SQUARE(muRA) + SQUARE(muDec) > 5.0*SQRT(
        SQUARE(muRA*sigMuRA)+SQUARE(muDec*sigMuDec)
       )
/* UKIDSS DR2 rows returned: 1 (count=0)
   Execution time:           01m 00s  */
\end{verbatim}

\noindent
At the time of writing no second epoch observations have
been taken for the UKIDSS LAS, so this query returns zero
in releases up to and including DR2.

For the count of unpaired objects, use of the view
\verb+lasReliablePointSource+ is recommended. Examination of the
available table attributes in the view (see Section~\ref{browser};
the attribute list is the same as the base table \verb+lasSource+
from which the view is derived) shows first-- and second--epoch
attribute names are prefixed by~\verb+j_1+ and~\verb+j_2+
respectively. Default values in one or other of the detection
unique identifiers \verb+ObjID+ for a given passband indicate
no merged pair in that band, so a count of unpaired sources is
simply obtained via

\begin{verbatim}
SELECT COUNT(*)
FROM   reliableLasPointSource
WHERE  
/* Specify detection at one epoch only:  */
       (j_1ObjID > 0 AND j_2ObjID < 0) OR
       (j_1ObjID < 0 AND j_2ObjID > 0)
/* UKIDSS DR2 rows returned: 1 (count=827968)
   Execution time:           01m 06s  */
\end{verbatim}

\noindent
where the test condition is for a default detection identifier
value (i.e.~no detection) at one or other, but not both,
of the two epochs. Once again, because no second epoch 
observations are available presently, this query simply
returns a count of all objects in the view since the
definition of \verb+reliableLasPointSource+ excludes
any object not detected at \verb+j_1+.

\subsubsection{Deep galaxy catalogues}
\label{U5}

Usage example~U5 in Appendix~\ref{usages} concerns
user--selected galaxy catalogues. The following simple SQL
example shows how to do this for the UKIDSS DXS:

\begin{verbatim}
SELECT ra, dec, 
/* De-reddened Petrosian magnitude and 
   fixed aperture colour: */
       jPetroMag-aj as j, 
       (jAperMag3-aj)-(kAperMag3-ak) as jmk
FROM   reliableDxsSource
WHERE  
/* Classification cut to exclude all point 
   sources: */
       mergedClass NOT BETWEEN -1 AND 0 AND
/* Exclude any sources with poorly or undefined 
   Petrosian mags: */
       jPetroMagErr BETWEEN 0 AND 0.2 AND
       kPetroMagErr BETWEEN 0 AND 0.2
/* UKIDSS DR2 rows returned: 142,996
   Execution time:           00m 06s  */
\end{verbatim}

\begin{figure}
\centerline{\includegraphics[width=84mm,angle=270]{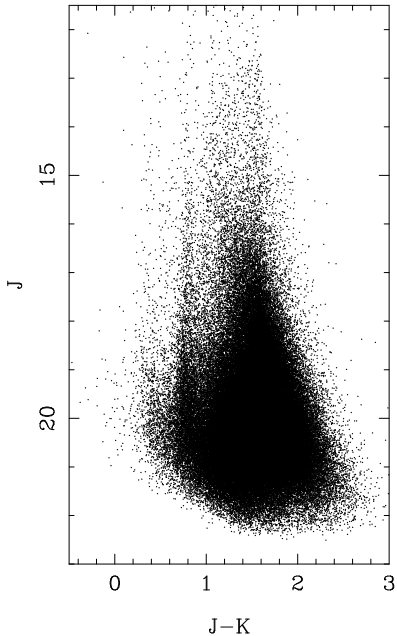}}
\caption{Colour--magnitude diagram in J versus J$-$K showing the results of the
query in Section~\ref{U5}.}
\label{dxscmd}
\end{figure}

\noindent
Here, we use the available view \verb+reliableDxsSource+ to
define a clean (but necessarily incomplete) selection,
excluding point--like sources. Several choices are available
as regards extended source flux measures -- see the entry for
the base table \verb+dxsSource+ in the schema browser
(described in Section~\ref{browser}). In this case, we have
chosen the Petrosian apparent magnitude, dereddened for
foreground Galactic extinction (Schlegel, Finkbeiner \& Davis~1998;
Bonifacio, Monai \& Beers~2000) and fixed 2\arcsec\ diameter
apertures for a colour index. A colour--magnitude diagram
is shown in Figure~\ref{dxscmd}.

\noindent
The spatial extent of the deep stacked UKIDSS surveys is easily determined
in SQL by a number of methods. The simplest is illustrated for the UDS
as follows:

\begin{verbatim}
SELECT MIN(ra),MAX(ra),MIN(dec),MAX(dec), (
        (MAX(ra)-MIN(ra))*COS(RADIANS(AVG(dec))))*
        (MAX(dec)-MIN(dec)
       ) AS area
FROM   udsSource
/* UKIDSS DR2 rows returned: 1
   Execution time:           00m 03s  */
\end{verbatim}

\noindent
This query returns the extent of the UDS in~RA and~Dec and the area
covered: 0.89~square degrees (more sophisticated examples concerning
areal coverage information are given in Section~\ref{U6}). As a
further example of galaxy catalogue selection, consider the following
query:

\newpage
\begin{verbatim}
SELECT CAST(ROUND(kab*2.0,0) AS INT)/2.0 AS K_AB, 
       LOG10(COUNT(*)/0.89) AS logN
FROM   (
       SELECT (kPetroMag-ak)+1.900 AS kab
       FROM   udsSource
       WHERE  mergedClass NOT BETWEEN -1 AND 0 AND
              jPetroMag > 0.0 AND
              kPetroMag > 0.0 
       ) AS T
GROUP BY CAST(ROUND(kab*2.0,0) AS INT)/2.0
ORDER BY CAST(ROUND(kab*2.0,0) AS INT)/2.0
/* UKIDSS DR2 rows returned: 42
   Execution time:           00m 03s  */
\end{verbatim}

\begin{figure}
\centerline{\includegraphics[width=55mm,angle=270]{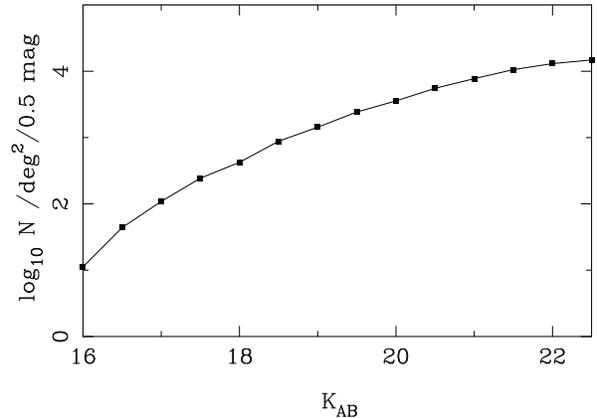}}
\caption{Galaxy number--magnitude counts in the UDS from the final 
query in Section~\ref{U5} (cf.~Figure~2 of Lane et al.~2007).}
\label{udsnm}
\end{figure}

\noindent
This consists of a nested subquery to select UDS galaxy catalogue
K$_{\rm AB}$ magnitudes via some simple predicates (note that
Vega--to--AB magnitude conversion constants are provided for
each WFCAM passband in table \verb+Filter+). The outer query
uses a combination of SQL functions to bin up counts of the
number of galaxies in 0.5 magnitude bins via grouping 
(see Appendix~\ref{sql}) within the appropriate ranges. The
results are plotted in Figure~\ref{udsnm}, and are in
good agreements with similar counts in Figure~2 of
Lane et al.~(2007) at the faint end where galaxies dominate
over stars in the counts.

\subsubsection{Star counts in cells in the UKIDSS GPS}
\label{U8}

One of the (many) advantages to the availability of a flexible SQL
interface in the WSA is that it allows the user to make summaries of
the data held without recourse to downloading entire source catalogues.
For example, in the UKIDSS DR2 the GPS merged source table contains
$3.6\times10^8$ rows; with a row length of $\sim1$~kilobyte the DR2
GPS merged source catalogue is over one third of a terabyte in size.
Usage example~U8 in Appendix~\ref{usages} shows a typical example where
star counts in cells (in this case in spherical polar co-ordinate
space) are required as a broad--brush summary of the catalogue. SQL
provides several functions that make counts in bins in arbitrary 
parameter space relatively straightforward:

\begin{verbatim}
SELECT   CAST(ROUND(l*6.0,0) AS INT)/6.0 AS lon, 
         CAST(ROUND(b*6.0,0) AS INT)/6.0 AS lat,
         COUNT(*)                        AS num
FROM     gpsSource
WHERE    k_1Class BETWEEN -2 AND -1 AND 
         k_1ppErrBits < 256         AND
/* Make a seamless selection (i.e. exclude 
   duplicates) in any overlap regions: */
         (priOrSec=0 OR priOrSec=frameSetID)
/* Bin up in 10 arcmin x 10 arcmin cells: */
GROUP BY CAST(ROUND(l*6.0,0) AS INT)/6.0,
         CAST(ROUND(b*6.0,0) AS INT)/6.0
/* UKIDSS DR2 rows returned: 28,026
   Execution time:           72m 00s  */
\end{verbatim}

\begin{figure*}
\centerline{\includegraphics[width=35mm,angle=270]{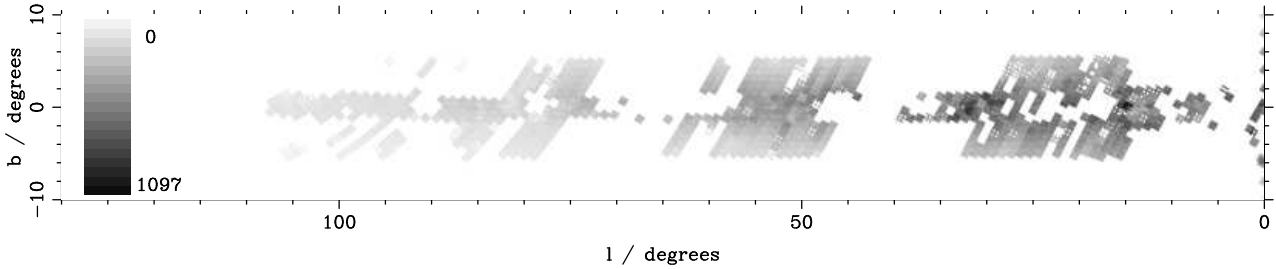}}
\caption{K--band star counts in the UKIDSS GPS produced by the query given
in Section~\ref{U8}. The scale bar is in units of stars per square 
arcminute.}
\label{gpscounts}
\end{figure*}

\noindent
In this example, nearest integer values of $l\times6$ and
$b\times6$, where $l,b$ (in units of degrees) are Galactic longitude
and latitude respectively, yield cells of size $10\times10$~arcmin$^2$.
We have chosen to use K--band star counts in this case, since this
passband has the most GPS data at DR2. Note the use of the predicate
\verb+(priOrSec=0 OR priOrSec=frameSetID)+. This uses the `primary or
secondary' flag attribute to select only those sources that have no
duplicates (\verb+priOrSec+=0) or primaries in the presence of duplicates
(\verb+priOrSec+ points to the current frame set identifier, 
indicating the source is duplicated
but that the current record is the best one to use); conversely, all the
secondaries of duplicates (and only those secondaries) could be selected 
by specifying \verb+priOrSec>0 AND priOrSec<>frameSetID+.
The results of the seamless selection in the query above are shown in 
Figure~\ref{gpscounts}.

\subsubsection{Optical/infrared selection of QSO candidates}
\label{U2}

Usage~U2 in Appendix~\ref{usages}
requires two selections: (i) a set of point sources
satisfying certain optical/infrared colour cuts; and (ii) a~1~in $10^4$
sampling of all point sources without those colour cuts. Both are easily
achieved in SQL -- the availability of the view \verb+lasPointSource+
is particularly convenient. The following query provides selection~(i):

\begin{verbatim}
SELECT psfMag_i-psfMag_z    AS imz,
       psfMag_z-j_1AperMag3 AS zmj,
       psfMag_i-yAperMag3   AS imy,
       ymj_1Pnt             AS ymj
FROM   lasPointSource        AS s,
       lasSourceXDR5PhotoObj AS x,
       BestDR5..PhotoObj     AS p
WHERE
/* Join predicates: */
       s.sourceID   = x.masterObjID AND
       x.slaveObjID = p.objID       AND
       x.distanceMins < 1.0/60.0    AND
/* Select only the nearest primary SDSS 
   point source crossmatch: */
       x.distanceMins IN (
         SELECT MIN(distanceMins)
         FROM   lasSourceXDR5PhotoObj 
         WHERE  masterObjID = x.masterObjID AND
                sdssPrimary = 1             AND
                sdssType    = 6
       ) AND
/* Remove any default SDSS mags: */
       psfMag_i > 0.0 AND
/* Colour cuts for high-z QSOs from 
   Hewett et al. (2006) and Venemans
   et al. (2007): */
       psfMag_i-yAperMag3 > 4.0 AND
       ymj_1Pnt           < 0.8 AND
       psfMagErr_u > 0.3 AND
       psfMagErr_g > 0.3 AND
       psfMagErr_r > 0.3 
/* UKIDSS DR2 rows returned: 12
   Execution time:           19m 56s  */
\end{verbatim}

\noindent
where the colour cuts have been determined with reference to Hewitt
et al.~(2006) and Venemans et al.~(2007).
In fact, usage example U2 is somewhat unrealistic in 
that the `legacy' SDSS lacks the depth to detect QSOs having $z\sim7$ as
illustrated in Venemans et al.~(2007); optical drop--out techniques
(see later) or deeper optical data are needed for the most highly
redshifted QSOs. Furthermore, some contamination
from differently deblended sources and poorly photometered sources near
very bright stars is present in exactly the position where the high
redshift QSO locus is expected to lie. However the SQL provided here 
serves at least to illustrate how to ask this kind of question in the 
WSA; moreover, it produces a list of a dozen candidates which is a viable
number for closer scrutiny (e.g.\ inspection of image thumbnails and
subsequent spectroscopic follow--up).

\begin{figure}
\centerline{\includegraphics[width=84mm,angle=270]{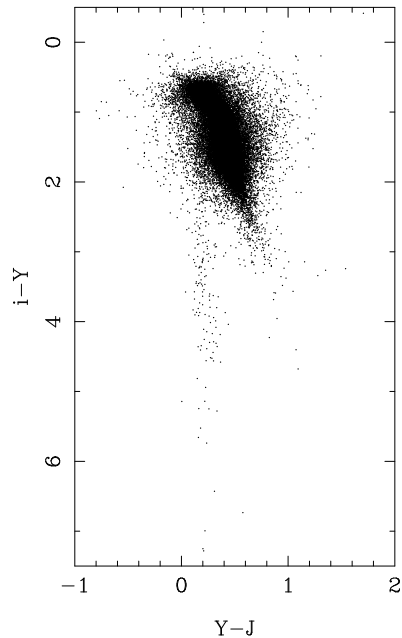}}
\caption{Two--colour diagram (cf.~Figure~5 of Hewitt et al.~2006)
illustrating the principal colour space of optical/infrared QSO candidate
selection (see Section~\ref{U2}).}
\label{qso2col}
\end{figure}

For selection (ii), removing the colour cut predicates and adding the
predicate \verb+... AND (sourceID%10000)=0+ will select one in every
$10^4$ sources randomly scattered over the survey area (the 
``\verb+%+'' modulo operator returns the remainder of the number on 
the left after dividing by that on the right). This is because
\verb+sourceID+ is assigned sequentially in the source merging
procedure and for large increments this attribute is not strongly
correlated with position. The results are illustrated in the
two--colour diagram in Figure~\ref{qso2col} (1 in~10 sources
plotted from UKIDSS DR2 cross--matched with SDSS DR5 rather than
a~1--in--$10^4$ sampling).

\subsubsection{A wide--area, shallow galaxy catalogue}
\label{U6}

Usage example~U6 in Appendix~\ref{usages} specifies the selection of
a galaxy catalogue with full optical/infrared photometry
to K=18.4 from the intersection of the UKIDSS LAS
and SDSS optical survey. In the following example, we further extend 
this usage mode to extract redshift information from SDSS 
spectroscopy, and compute absolute magnitudes M$_{\rm K}$ via an
Einstein--de Sitter cosmological distance modulus with Hubble
constant $H_0=75$~km~s$^{-1}$Mpc$^{-1}$, all in SQL.
The nearest crossmatch between the LAS and SDSS with a
matching tolerance of 2\arcsec\ is selected:

\begin{verbatim}
SELECT s.ra as alpha, s.dec as delta,
/* Remove Galactic foreground reddening: */
       (petroMag_u-extinction_u)     AS u, 
       (petroMag_g-extinction_g)     AS g, 
       (petroMag_r-extinction_r)     AS r, 
       (petroMag_i-extinction_i)     AS i, 
       (petroMag_z-extinction_z)     AS z,
       (yPetroMag-ay)                AS y, 
       (j_1PetroMag-aj)              AS j,
       (hPetroMag-ah)                AS h,
       (kPetroMag-ak)                AS k,
       z.z                           AS redshift,
       (modelMag_g-extinction_g) -
        (modelMag_r-extinction_r)    AS gmr,
       (yAperMag3-ay)-(kAperMag3-ak) AS ymk,
       (modelMag_u-extinction_u) - 
        (modelMag_g-extinction_g)    AS umg,       
/* Einstein-de Sitter cosmology distance modulus
   (note no K-correction, no evolution correction, 
    and no internal extinction):  */
       (kPetroMag-ak) - 25 - 5*(
        LOG10(2*2.998e5*(1+z.z-SQRT(1+z.z))/75)
       ) AS M_K
FROM   lasExtendedSource AS s, 
       lasSourceXDR5PhotoObj AS x, 
       BestDR5..PhotoObj AS p, 
       BestDR5..SpecObj AS z
WHERE
/* Join criteria:   */
       z.specObjID=p.specObjID    AND
       s.sourceID = x.masterObjID AND
       p.objID = x.slaveObjID     AND
       x.distanceMins IN (
       SELECT MIN(distanceMins)
       FROM   lasSourceXDR5PhotoObj
       WHERE  masterObjID = x.masterObjID AND
              distanceMins < 2.0/60.0
       ) AND
/* Dereddened magnitude cut as specified:  */
       (kPetroMag-ak) BETWEEN 0.0 AND 18.4 AND
       yPetroMag > 0  AND
       modelMag_u > 0 AND
       modelMag_g > 0 AND
       modelMag_r > 0 AND
/* Exclude any non spectroscopic redshift 
   objects for a clean sample: */
       z.z BETWEEN 0.01 AND 0.15 AND
       z.zWarning=0
/* UKIDSS DR2 rows returned: 8,086
   Execution time:           05m 13s  */
\end{verbatim}

\begin{figure}
\centerline{\includegraphics[width=55mm,angle=270]{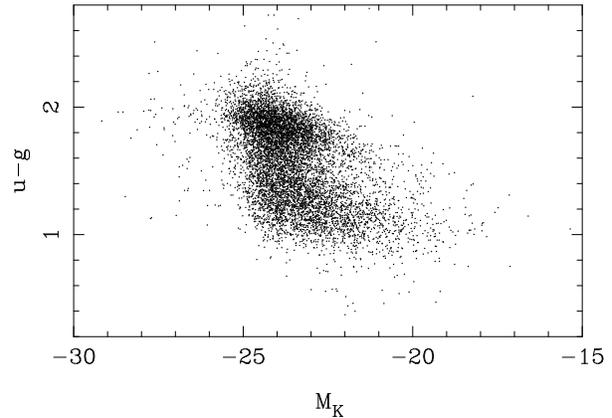}}
\caption{Absolute--magnitude versus colour plot for a wide--area,
shallow galaxy catalogue extracted using the query in 
Section~\ref{U6} which trawls the UKIDSS LAS--DR2 and SDSS--DR5
crossmatch (see text).}
\label{laszgals}
\end{figure}

\noindent
In Figure~\ref{laszgals} we plot M$_{\rm K}$ (as a proxy for total
stellar mass) versus (u$-$g) which shows a bright red clump of ellipticals
along with a sequence of fainter, bluer star--forming and/or spiral
galaxies and finally yet fainter, bluer dwarfs.

\begin{figure*}
\centerline{\includegraphics[width=25mm,angle=270]{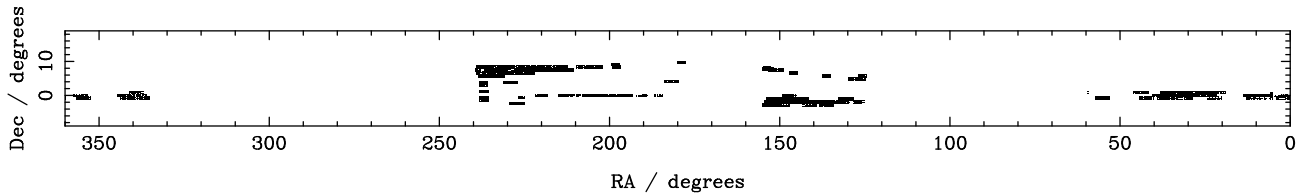}}
\caption{Areal coverage of the galaxy catalogue selection
described in Section~\ref{U6}.}
\label{lasarea}
\end{figure*}

Spatial sampling of selections from base table \verb+lasSource+
(or indeed any merged source table in the WSA) can be determined
in several ways. The simplest method (e.g.~for making an areal
coverage plot) is to use the central positions of the frame
sets available in \verb+lasMergeLog+:

\begin{verbatim}
SELECT ra,dec
FROM   lasMergeLog
WHERE  j_1mfID > 0 AND
       hmfID   > 0 AND
       kmfID   > 0
/* UKIDSS DR2 rows returned: 6,242
   Execution time:           00m 04s  */
\end{verbatim}

\noindent
where the predicates require coverage in passbands JHK, but
not necessarily~Y, as is the case in view \verb+lasExtendedSource+
for example. The size of each frame set in the LAS is the size of one
WFCAM detector, or 13.65~arcmin. Figure~\ref{lasarea} shows the area
covered by plotting squares of this size at
each co-ordinate pair returned by the query.

More sophisticated functionality is provided via use of Hierarchical
Triangular Mesh (HTM; Kunszt et al.~2000) indices which are available
as attribute \verb+htmID+ where an equatorial RA,Dec pair are
present in most WSA tables. For example, the set of HTM triangles
covering a given selection to a given HTM level (see Kunszt et al.~2000)
can be obtained using the SQL \verb+DISTINCT+ function along with
division by an integer power $N$ of 4 to mask to the $(20-N)^{\rm TH}$
level where the WSA uses 20--level indexing by default:

\begin{verbatim}
SELECT DISTINCT(htmID/POWER(4,20-12)) 
FROM   lasSource 
WHERE  ...
\end{verbatim}

\noindent
will return the identifiers of the HTM triangles at level 12
(areas\footnote{\tt http://www.sdss.jhu.edu/htm} between~0.86
and~1.8 square arcmin) covered by the LAS merged source table
for the given predicates. Libraries of various routines for the 
manipulation and translation of spatial co-ordinates an associated
HTM indices are are available online$^7$. Note that the areal
coverage maps returned by any of these queries are not the maps
of survey depth that would be needed to compute survey volume
corrections.

\subsubsection{Infrared colour--selected sources that are optical
drop--outs}
\label{tbds}

Usage~U1 in Appendix~\ref{usages} requires non--detection in optical (iz)
passbands for an infrared colour--selected sample of point sources as
cool, substellar candidates (see, for example, Kendall et al.~2007).
One could envisage this being achieved within the archive by
automatically placing apertures in optical images (i.e.~SDSS pixel
data) at positions having infrared detections. In fact, it is much
simpler to use the cross--neighbour functionality, requiring 
non--detection in the optical above a certain limit within a given
radius of an infrared source. In this way it is possible to make
a manageable candidate list in a single SQL 
\verb+SELECT+ statement. In~U1, it is envisioned that the user 
develops the query for a rare object search by refining the search
predicates. The starting point would be as follows:

\begin{verbatim}
SELECT COUNT(*)
FROM   lasSource
WHERE
/* Colour cuts for mid-T & later: */
       ymj_1Pnt > 0.5 AND
       j_1mhPnt < 0.0 AND
/* Source not detected above 2sigma within  
   1" in SDSS-DR5 i' or z':                */
       sourceID NOT IN (
       SELECT masterObjID
       FROM   lasSourceXDR5PhotoObj AS x,
              BestDR5..PhotoObj     AS p
       WHERE  p.objID = x.slaveObjID   AND
              (psfMagErr_i < 0.5       OR 
               psfMagErr_z < 0.5)      AND
              x.distanceMins < 1.0/60.0
        ) AND
/* Use only frame sets overlapping with 
   SDSS-DR5:                            */
       frameSetID IN (
       SELECT DISTINCT(frameSetID)
       FROM   lasSource             AS s,
              lasSourceXDR5PhotoObj AS x
       WHERE  s.sourceID = x.masterObjID
       )
/* UKIDSS DR2 rows returned: 1 (count=46,141)
   Execution time:           16m 52s  */
\end{verbatim} 

\noindent
which counts 46,141 candidates in the UKIDSS DR2 cross--match with SDSS--DR5.
The first two predicates simply apply a colour cut based on prior knowledge
of the objects being sought. The third predicate involves a subquery to
exclude any object that has an optical counterpart above the specified
limit ($2\sigma$) in the SDSS, and a further nested
subquery to only use the {\em nearest} optical cross--match within
1\arcsec. Finally, a predicate subquery
specifies that only LAS frame sets that contain SDSS cross--matches
should be used in this search, since if any LAS imaging data are
outwith the area covered by the SDSS, they must be excluded since all
infrared sources in those regions would be counted as optical 
non--detections.

Clearly, $\sim4.6\times10^4$~candidates is an impractically large list 
for any useful purpose. The predicates need to be expanded to reduce 
the list of unwanted and spurious sources prior to a more intensive 
inspection of image thumbnails or indeed spectroscopic follow--up on 
large aperture facilities. Addition of the following predicates:

\begin{verbatim}
/* Unduplicated or primary duplicates only: */
        (priOrSec = 0 OR priOrSec = frameSetID) AND
/* Generally good quality: */
        yppErrBits   < 256 AND
        j_1ppErrBits < 256 AND
        hppErrBits   < 256 AND
/* Point-like morphological classification: */
        mergedClass=-1 AND
        mergedClassStat BETWEEN -3.0 AND +3.0 AND
/* Reasonably circular images in YJH: */
        yEll   < 0.35 AND 
        j_1Ell < 0.35 AND 
        hEll   < 0.35 AND
/* IR pairs within 0.5 arcsec: */
        j_1Xi  BETWEEN -0.5 AND +0.5 AND
        j_1Eta BETWEEN -0.5 AND +0.5 AND
        hXi    BETWEEN -0.5 AND +0.5 AND
        hEta   BETWEEN -0.5 AND +0.5 AND
/* YJ measured to 5 sigma and H to 4sigma: */
        yAperMag3Err   < 0.20 AND
        j_1AperMag3Err < 0.20 AND
        hAperMag3Err   < 0.25
/* UKIDSS DR2 rows returned: 1 (count=25)
   Execution time:           00m 48s  */
\end{verbatim}

\noindent
reduces the number of candidates to~25. The first predicate 
limits the search
to unique objects where duplicates exist in overlap regions; the next set
makes cuts on quality error flags; the next limits the search to point--like,
circular sources making generous allowance for noisy, high ellipticities at
low signal--to--noise; the penultimate predicate set restricts the 
selection to YJH pairs within 0.5\arcsec\ boxes 
(the LAS pairing criterion is a generous 2.0\arcsec\ -- 
e.g.~Table~\ref{pairtols}). All the predicates on attributes common to
all passbands are applied
across the relevant filter passbands (YJH) to limit the sample selection
to those sources appearing in all three. Substituting 
\verb+SELECT COUNT(*)+ with

\begin{verbatim}
SELECT dbo.fIAUnameLAS(ra,dec),
       yAperMag3, 
       ymj_1Pnt,ymj_1PntErr,
       j_1mhPnt,j_1mhPntErr
\end{verbatim}

\noindent
and including \verb+ORDER BY ra+ at the end of the query yields the
results shown in Table~\ref{Ttypes}. Note the syntax and use of the 
function \verb+fIAUnameLAS()+ to automatically output IAU standard
names for any target. The candidate sample produced
by the full query includes spectroscopically confirmed T~dwarfs 
discussed in Lodieu et al.~(2007b) and references therein.

\begin{table*}
\caption{Candidate T--type brown dwarfs extracted from the UKIDSS LAS
at Data Release~2 using the example query given in Section~\ref{tbds}.
The final column indicates the assigned spectral type (if known) from
follow--up observations as reported in Warren et al.~(2007b) for ULAS~J0034
and Lodieu et al.~(2007b) otherwise.}
\label{Ttypes}
\begin{tabular}{lrrrrrc}
\hline
\multicolumn{1}{c}{Candidate name} &
  \multicolumn{1}{c}{Y}  & 
  \multicolumn{1}{c}{Y$-$J}   & 
  \multicolumn{1}{c}{$\sigma_{\rm Y-J}$}  & 
  \multicolumn{1}{c}{J$-$H}   & 
  \multicolumn{1}{c}{$\sigma_{\rm J-H}$}  & 
  \multicolumn{1}{c}{Sp. type} \\
\hline
ULAS J002422.93+002247.9 	& 	19.604 	& 	1.441 	& 	0.170 	& 	$-$0.093 	& 	0.174 	& T4.5\\
ULAS J003402.77--005206.7 	& 	18.905 	& 	0.769 	& 	0.130 	& 	$-$0.398 	& 	0.234 	& T8.5\\
ULAS J013738.55+011808.0 	& 	19.497 	& 	0.762 	& 	0.168 	& 	$-$0.013 	& 	0.231 	& \\
ULAS J020331.51+012350.8 	& 	19.276 	& 	0.663 	& 	0.157 	& 	$-$0.001 	& 	0.219 	& \\
ULAS J020336.94--010231.1 	& 	19.056 	& 	1.014 	& 	0.109 	& 	$-$0.284 	& 	0.132 	& T5.0\\
ULAS J021127.17+011606.1 	& 	19.226 	& 	0.647 	& 	0.124 	& 	$-$0.016 	& 	0.198 	& \\
ULAS J022200.43--002410.5 	& 	19.734 	& 	1.084 	& 	0.151 	& 	$-$0.033 	& 	0.164 	& \\
ULAS J024641.36+011800.5 	& 	19.767 	& 	0.615 	& 	0.203 	& 	$-$0.072 	& 	0.271 	& \\
ULAS J024642.92+011628.9 	& 	19.685 	& 	0.880 	& 	0.169 	& 	$-$0.194 	& 	0.214 	& \\
ULAS J025321.88+012319.8 	& 	19.961 	& 	0.726 	& 	0.192 	& 	$-$0.052 	& 	0.256 	& \\
ULAS J030135.17+011903.8 	& 	19.524 	& 	0.685 	& 	0.165 	& 	$-$0.160 	& 	0.254 	& \\
ULAS J083554.96--012556.6 	& 	18.942 	& 	0.695 	& 	0.110 	& 	$-$0.123 	& 	0.189 	& \\
ULAS J084719.65--012701.2 	& 	19.206 	& 	0.551 	& 	0.146 	& 	$-$0.052 	& 	0.250 	& \\
ULAS J093956.86--012109.7 	& 	19.327 	& 	0.638 	& 	0.126 	& 	$-$0.154 	& 	0.249 	& \\
ULAS J094806.06+064805.0 	& 	20.030 	& 	1.177 	& 	0.158 	& 	$-$0.709 	& 	0.230 	& T7.0\\
ULAS J100513.82--003704.2 	& 	19.660 	& 	0.758 	& 	0.131 	& 	$-$0.004 	& 	0.228 	& \\
ULAS J100759.90--010031.1 	& 	19.818 	& 	1.146 	& 	0.136 	& 	$-$0.119 	& 	0.195 	& T5.5\\
ULAS J115038.79+094942.8 	& 	19.588 	& 	1.012 	& 	0.139 	& 	$-$0.751 	& 	0.222 	& \\
ULAS J115811.10+094236.1 	& 	19.899 	& 	0.550 	& 	0.186 	& 	$-$0.003 	& 	0.226 	& \\
ULAS J131346.93+075451.8 	& 	20.238 	& 	0.992 	& 	0.196 	& 	$-$0.189 	& 	0.249 	& \\
ULAS J145514.50+061655.8 	& 	20.163 	& 	0.784 	& 	0.229 	& 	$-$0.006 	& 	0.257 	& \\
ULAS J150508.56+061547.1 	& 	19.871 	& 	0.690 	& 	0.161 	& 	$-$0.080 	& 	0.268 	& \\
ULAS J155120.90+075159.5 	& 	19.121 	& 	0.539 	& 	0.104 	& 	$-$0.182 	& 	0.234 	& \\
ULAS J223132.02+012334.5 	& 	19.519 	& 	0.517 	& 	0.185 	& 	$-$0.095 	& 	0.272 	& \\
ULAS J224238.02+011804.3 	& 	19.499 	& 	0.547 	& 	0.174 	& 	$-$0.013 	& 	0.259 	& \\
\hline
\end{tabular}
\end{table*}

\section{Conclusion}
\label{concs}

We have described the WFCAM Science archive (WSA), which is the end point
in the data flow of UKIRT WFCAM data in the VISTA Data Flow System,
and the primary point of access for users of survey science products,
especially those of the United Kingdom Infrared Deep Sky Survey (UKIDSS). 
In particular, we have described:
\begin{itemize}
\item how the top--level requirements and typical usage
modes informed the design of the WSA;
\item the arrangement of survey data in terms of a set of related tables;
\item the implementation of the archive within a commercial
relational database management system;
\item the curation procedures employed to create science--ready
survey catalogues from standard pipeline--processed products;
\item example real--world astronomy usage modes along with typical results.
\end{itemize}

The WSA is the prototype science archive for the VISTA surveys, and the
design of the VISTA Science Archive will follow closely that described
here.

%++++++++++ ACKNOWLEDGEMENTS ++++++++++++++++++++++++++

\section{Acknowledgements}  \label{acknow}

Many people have contributed to the development of the WFCAM Science
Archive. We would like to thank especially our colleagues in the
Sloan Digital Sky Survey science archive project: Alex Szalay,
Jan Vandenberg, Maria Nieto--Santisteban, Peter Kunszt, Tamas Budav\'{a}ri,
and Ani Thakar at Johns Hopkins University, and
Jim Gray from Microsoft Research, for all their advice during the
design and implementation stages. We thank Microsoft also for
providing software via the Microsoft Development Network Academic
Alliance, and eGenix for providing a free, non--commercial licence
for Python middleware. We would like to acknowledge help and
support from the following individuals: Luc Simard, Frossie Economu,
John Lightfoot, Tim Jenness, Andy Vick, Dave Lunney, Mark Casali, Mike Irwin,
Jim Lewis, Peter Bunclark,
Richard McMahon, Martin Hill, John Taylor, Mark Holliman,
Jonathan Dearden, and Horst Meyerdierks.
Eclipse Computing of Ayrshire have been unfailing in their supply of
test equipment and final hardware solutions at highly competitive
prices. The WSA team would like to acknowledge much helpful interaction
with University of Edinburgh staff in the Schools of Physics and Informatics
and the National e--Science Centre, and advice from Andy Knox of IBM
concerning scalable hardware solutions. Finally, we would like to extend
our thanks to Ian Carney of the Oracle Corporation for introducing us
to the delights of entity--relationship modelling.

Financial resources for VISTA/WFCAM Science Archive development were 
provided by the UK Science and Technology Facilities Council
(STFC; formerly the Particle Physics and Astronomy Research Council)
via e--Science project funding for the VISTA Data Flow 
System. Operations resources, including provision for hardware, are provided
in the main by the STFC, but we would also like to acknowledge 
significant contributions to hardware funding from University of Edinburgh 
sources.

\vspace*{12pt}
This paper is dedicated to our colleague and friend Jim Gray.

%%%%%%%%%%%%%%%%%%%%%%%%%%%%%%%%%%%%%%%%%%%%%%%%%%%%%%%%%%%%%%%%%%%%%%%%
%    REFERENCES
%%%%%%%%%%%%%%%%%%%%%%%%%%%%%%%%%%%%%%%%%%%%%%%%%%%%%%%%%%%%%%%%%%%%%%%%%

%\newpage

\appendix

\section{Twenty usages of the WFCAM Science Archive}
\label{usages}

Here we reproduce the typical usage modes of the WSA that were
developed in collaboration with the UKIDSS user community during the
design phase of the project (more details are available 
online\footnote{\tt http://surveys.roe.ac.uk/wsa/pubs/wsausage.html}):

{\bf U1}:  Count the number of sources in the LAS which satisfy the colour constraints (Y--J)~$>1.0$, (J--H)~$<0.5$ where SDSS iz flux limits at the same position are less than 2$\sigma$. User then refines the query as necessary to give a reasonable number of candidates. When satisfied, the user requests a list, selecting output attributes from those available for the LAS, and finder charts in JHK for each object.

{\bf U2}:  List all star-like objects with izYJHK SDSS/UKIDSS--LAS colours consistent with the colours of quasars at redshifts $5.8 < z < 7.2$ or $z > 7.2$ (user specifies cuts in colour space). Return plots of (i--z) versus (z--J) and (i--Y) versus (Y--J) with these sources plotted in a specified symbol type, with 1 in every 10,000 other stellar sources plotted as points.

{\bf U3}:  For a given cluster target in the UKIDSS GCS, make a candidate membership list via selection of stellar sources in colour--magnitude, colour--colour and proper motion space. Cross--correlate the candidate list against a user--supplied catalogue of optical/near--infrared detections in the same region.

{\bf U4}:  From the UKIDSS LAS, provide a list of all stellar objects that have measured proper motions greater than 5x their estimated proper motion error; additionally give a count of all stellar objects that are unpaired between the two epochs of the LAS observations with specified conditions on image quality flags. User then refines these conditions to produce a manageable list of very high proper motion candidate stars. Return finder charts in JHK for all candidates.

{\bf U5}:  From the UKIDSS DXS \& UDS, construct galaxy catalogues. User selects all non-stellar sources satisfying quality criteria. User also requires the spatial sampling of this catalogue. Cross-correlate the galaxy catalogues against user-supplied optical catalogue in the same region.

{\bf U6}:  From the UKIDSS LAS, construct a galaxy catalogue for all non-stellar sources satisfying K~$<18.4$ and given quality criteria; return full photometric list from SDSS \& UKIDSS: ugrizYJHK. User also requires the spatial sampling of this catalogue.

{\bf U7}:  From the UKIDSS UDS, select a sample of galaxies with colours and morphology consistent with being elliptical galaxies. Provide a spatial mask to enable determination of sample characteristics. Provide a measure of the half-light radius for each galaxy.

{\bf U8}:  From the UKIDSS GPS, provide star counts in 10 arcmin cells on a grid in Galactic longitude and latitude; also provide a list of cells where there is any quality issue rendering that cell's value inaccurate.

{\bf U9}:  From the UKIDSS GPS, provide a list of all sources that have brightened by a given amount in the K band.

{\bf U10}:  Provide a plot of g--J vs J--K for all point--like sources detected in the UKIDSS/LAS survey subject to quality constraints. User interacts with the plot to fit a straight line (g--J)=a+b(J--K) to the main sequence stars. Then find all UKIDSS/LAS sources with g-J$>$a+b(J-K), 4$>$g--J$>$--1, and 3$>$J--K$>$--1.

{\bf U11}:  Construct H2--K difference image maps for all frames within a specified subregion surveyed by the GPS.

{\bf U12}:  Find all galaxies with a de Vaucouleurs profile and infrared colours consistent with being an elliptical galaxy in the Virgo region of the UKIDSS LAS.

{\bf U13}:  Given input co-ordinates and a search radius (arbitrary system and reference frame) provide a list of all WFCAM observations ever taken that contain data in all or part of the specified area.

{\bf U14}:  Provide a list of point-like sources with multiple epoch measurements which have light variations $>$~0.1 magnitudes in J, H or K.

{\bf U15}:  From any UKIDSS data, where multiple epoch measures exist for the same object, provide a list of anything moving more than X arcsec per hour.

{\bf U16}:  Provide a list of star--like objects that are 1\% rare for the 3--colour attributes.

{\bf U17}:  For a given device in a tile, give me all images from the UDS corresponding to that frame, stacked in 10 day bins.

{\bf U18}:  Give me a true colour JHK image mosaic using frames in the LAS centred at given co-ordinates (arbitrary reference frame and system) with 2 degree width and rebinned so that the entire mosaic is returned as a 2048x2048 pixel image.

{\bf U19}:  Find all detected sources from all UKIDSS surveys within 3x the error boxes of a user supplied list of X--ray transient sources.

{\bf U20}:  For all sources in a user--supplied radio catalogue of HII regions in the GPS, return the Br--gamma surface brightness in an aperture of X arcsec

\section{Structured Query Language data retrieval fundamentals}
\label{sql}

The basic, general form of a Structured Query Language 
(SQL; Klein \& Klein 2001) statement
for data retrieval (i.e.~a query statement) in an RDBMS is as follows:\\

\noindent
\verb+SELECT column-1 [, column-2, ...]+ \\
\verb+FROM   table-set-1 [, table-set-2, ...]+ \\
\verb+WHERE  condition-1 [ AND|OR condition-2 ... N]+ \hfill (B1) \\

The column definition is generally a comma--separated list of attribute
names from columns contained in the table set defined in the \verb+FROM+
clause, e.g.~\verb+SELECT ra, dec, frameSetID ...+, but great flexibility
is available in SQL: expressions involving literal constants, mathematical
functions and statistical aggregates are all possible:\\

\noindent
\verb+SELECT 'hello world'+ \hfill (B2) \\
\verb+SELECT ra/15.0 AS raHours, ...+ \hfill (B3) \\
\verb+SELECT AVG(COS(RADIANS(dec))), ...+ \hfill (B4) \\
\verb+SELECT COUNT(DISTINCT multiframeID), ...+ \hfill (B5) \\

\noindent
are all legal WSA SQL \verb+SELECT+ expressions; example (B2) is a complete
SQL statement that, insofar as SQL can be considered a programming language,
demonstrates the classic first step in learning the programming syntax --
it returns a results set consisting of one row having one column having the
specified string constant value. A more detailed explanation of SQL
\verb+SELECT+ is given 
online\footnote{http://surveys.roe.ac.uk/wsa/sqlcookbook.html}
at the WSA website in the `SQL Cookbook', while a complete description
including all standard clauses and non--standard Microsoft SQL~Server
extensions is available 
elsewhere\footnote{http://msdn2.microsoft.com/en-us/library/ms189826.aspx}.
Note that Microsoft SQL syntax is not case--sensitive -- mixed upper
and lower case is used in the examples in this paper for clarity only.

The table set definition in its simplest form consists of the name of a
single table, e.g.\\

\noindent
\verb+SELECT ra, dec, frameSetID+ \\
\verb+FROM   dxsMergeLog+ \hfill (B6) \\

\noindent
returns the equatorial co-ordinates of all frame sets in the UKIDSS DXS
along with their unique identification numbers that have been assigned
in the WSA curation procedure. Once again, great flexibility is
afforded in SQL in the table set definition: \verb+table-set-N+ may
be {\em any} expression that defines a tabular dataset, e.g.~a table
name, a view name, or even another \verb+SELECT+ statement. For
example, \\

\noindent
\verb+SELECT t.*+\\
\verb+FROM   (+\\
\verb+        SELECT ra, dec, frameSetID+\\
\verb+        FROM   dxsMergeLog+\\
\verb+       ) AS t+ \hfill (B7) \\

\noindent is an unnecessarily complicated, but nonetheless legal, SQL
equivalent to statement~B6 above (the nested \verb+SELECT+ is commonly
known as a subquery in this context; note the use of the alias ``t''
to conveniently label the subquery rowset for references elsewhere in
the statement).

By far the most common table set definition that a user will need when
retrieving data via free--form SQL statements is a comma--separated
list of related tables. With reference to the relational model in
Section~\ref{topLevel}, Figure~\ref{surveyProgERM}, consider the
case where a user wishes to obtain a list of the required filters
(WSA filter unique identifiers
\verb+filterID+ and number of multi--epoch passes in that filter)
for the UKIDSS programmes, along with generic information on each
programme. Since all the relevant information is
spread between tables \verb+Programme+ and \verb+RequiredFilters+, a
selection from those two is required:\\

\noindent 
\verb+SELECT t1.programmeID, t1.description, t2.*+\\
\verb+FROM   Programme AS t1, RequiredFilters AS t2+ \hfill (B8) \\

\noindent
This query, however, results in the {\em cartesian product} of the 
two tables rather than a union of associated rows. Most of the rows
in the results set produced by~B8 are of course meaningless, since
all~$N$ rows in \verb+Programme+ are joined, one by one, with 
all~$M$ rows of in \verb+RequiredFilters+ resulting 
in~$N\times M$ rows. In order to produce the selection required,
a \verb+WHERE+ clause must be used to associate the related rows,\\

\noindent
\verb+WHERE  t1.programmeID = t2.programmeID+ \hfill (B9) \\

\noindent 
since any rows where \verb+programmeID+ is different in the two
tables are not related. Generally speaking, when querying data
across $N$ tables there should be {\em at least} $N-1$ \verb+WHERE+
clause filters associating related attributes across the tables. 
The attributes to use in
filtering are easily determined using the {\em WSA schema browser}
(see Section~\ref{browser} in the main text). They are nearly always
indexed {\em primary keys} in the RDBMS implementation so are
highlighted and are at the top of each table's attribute list. 
Moreover a {\em foreign key} reference is noted at the top of each 
table definition for every many--to--one relationship in the
data model; referencing attributes are generally the ones to filter on
in the \verb+WHERE+ clause of a join query. Implicit join queries
are very common in {\em normalised} relational database
(e.g.~Section~\ref{queries} in the main text). The RDBMS design
is optimised for the normal form, required storage space is minimised,
and query performance is optimised for speed.

Otherwise, the \verb+WHERE+ clause is simply a list of conditional
statements linked by logical operators (usually \verb+AND+). These
conditions are known as {\em predicates}. Comparison
predicates are common:\\

\noindent
\verb@WHERE (ra/15.0 < 12.0 OR dec >=+35.0) AND@ 
\verb@      filterID <> 3@ \hfill (B10) \\

\noindent
Other types of predicate are defined in SQL -- again, see the WSA
SQL Cookbook or other online guides to the language.

Finally, there are some powerful optional clauses available to the
\verb+SELECT+ statement. An \verb+ORDER BY+ clause can be
specified, which sorts the results set returned by the specified
expression. For example,\\

\newpage
\noindent
\verb+SELECT ra, dec, frameSetID+\\
\verb+FROM   dxsMergeLog+\\
\verb+ORDER BY ra ASC+ \hfill (B11) \\

\noindent returns the same rows as~B6 but in order of increasing,
i.e.~ASCending, RA; specify DESC for descending order. Note that
without an \verb+ORDER BY+ clause, the order in which rows are retrieved
from an RDBMS is undefined and, moreover, generally unrepeatable -- the
order can change between two consecutive runs of the same query.

Furthermore, particularly useful for summarising the characteristics of
data in very large tables is the \verb+GROUP BY+ optional clause. This,
along with bulit--in aggregate functions, enables the user to produce
summary quantities or statistics for large amounts of data arbitrarily
grouped together on an expression involving one or more column names.
The \verb+GROUP BY+ clause is best illustrated with a few examples.\\

\noindent
\verb+SELECT   filterID, COUNT(*) AS totalFrames+\\
\verb+FROM     Multiframe+\\
\verb+GROUP BY filterID+ \hfill (B12) \\

\noindent
is a simple example which returns a count of the number of frames in
each of the different filters used in observing. Note the use of the
{\em aggregate function} \verb+COUNT+ in~B12 above. Queries involving
\verb+GROUP BY+ will generally use such built--in aggregate functions,
and this is a particularly powerful combination. Other aggregate
functions are available including minimum/maximum 
(\verb+MIN+/\verb+MAX+), average (\verb+AVG+), summation (\verb+SUM+)
and statistical aggregates, for example standard deviation
(\verb+STDEV+). A slightly more complicated example is\\

\noindent
\verb+SELECT   frameSetID, AVG(ra) AS meanRA,+\\
\verb+         AVG(dec) AS meanDec,+\\
\verb+         COUNT(*) AS numSources+\\
\verb+FROM     lasSource+\\
\verb+GROUP BY frameSetID+\\
\verb+HAVING   AVG(dec) > 0.0+ \hfill (B13) \\

\noindent
which returns a list of all northern hemisphere frame sets in the
UKIDSS LAS, their mean RAs/Decs and a count of the number of sources
in each. Note the additional \verb+HAVING+ clause: just as \verb+SELECT+
may have a \verb+WHERE+ clause to filter rows in the table(s) specified
in the \verb+FROM+ clause, \verb+GROUP BY+ may include a \verb+HAVING+
clause to filter rows in the table formed by the specified grouping.

%+++++++++ CLOSE OUT +++++++++++++++++++++++++++++++
\label{lastpage}

\end{document}